\documentclass[iop,numberedappendix]{emulateapj}
\usepackage{natbib}
\usepackage{graphicx}
\usepackage{rotating}
\usepackage{verbatim}

\slugcomment{Revised version November 11 2010 -- Accepted for publication in ApJ}

\shorttitle{THE BULK OF THE BLACK HOLE GROWTH SINCE $z\sim1$}
\shortauthors{CISTERNAS ET AL.}

\begin{document}

\title{The Bulk of the Black Hole Growth Since $\lowercase{z}\sim1$ Occurs in a Secular Universe:\\
    No Major Merger-AGN Connection$^{\star}$}


\author{Mauricio Cisternas$^{1,}$\altaffilmark{20},
 Knud Jahnke$^{1}$,
 Katherine J. Inskip$^{1}$,
 Jeyhan Kartaltepe$^{2}$,
 Anton M. Koekemoer$^{3}$,
 Thorsten Lisker$^{4}$,
 Aday R. Robaina$^{1,5}$,
 Marco Scodeggio$^{6}$,
 Kartik Sheth$^{7,8}$,
 Jonathan R. Trump$^{9}$,
 Ren\'{e} Andrae$^{1}$,
 Takamitsu Miyaji$^{10,11}$,
 Elisabeta Lusso$^{12}$,
 Marcella Brusa$^{13}$,
 Peter Capak$^{7}$,
 Nico Cappelluti$^{13}$,
 Francesca Civano$^{14}$,
 Olivier Ilbert$^{15}$,
 Chris D. Impey$^{9}$
 Alexie Leauthaud$^{16}$,
 Simon J. Lilly$^{17}$,
 Mara Salvato$^{18}$,
 Nick Z. Scoville$^{7}$,
 and Yoshi Taniguchi$^{19}$
 }
\email{cisternas@mpia.de}


\affil{$^{1}$ Max-Planck-Institut f\"{u}r Astronomie, K\"{o}nigstuhl 17, D-69117 Heidelberg, Germany}
\affil{$^{2}$ National Optical Astronomy Observatory, 950 North Cherry Avenue, Tucson, AZ 85 721, USA}
\affil{$^{3}$ Space Telescope Science Institute, 3700 San Martin Drive, Baltimore, MD 21218, USA}
\affil{$^{4}$ Astronomisches Rechen-Institut, Zentrum f\"{u}r Astronomie der Universit\"{a}t Heidelberg, M\"{o}nchhofstr. 12-14, 69120 Heidelberg, Germany}
\affil{$^{5}$ Instituto de Ciencias del Cosmos (ICC), Universidad de Barcelona-IEEC, Mart\'{\i} i Franqu\'{e}s 1, 08028 Barcelona, Spain}
\affil{$^{6}$ IASF-INAF, Via Bassini 15, I-20133 Milano, Italy}
\affil{$^{7}$ California Institute of Technology, 1200 East California Boulevard, MC 249-17, Pasadena, CA 91125, USA}
\affil{$^{8}$ Spitzer Space Center, California Institute of Technology, Pasadena, CA 91125, USA}
\affil{$^{9}$ Steward Observatory, University of Arizona, 933 North Cherry Avenue, Tucson, AZ 85721, USA}
\affil{$^{10}$ Instituto de Astronom\'ia, Universidad Nacional Aut\'onoma de M\'exico, Ensenada, M\'exico (PO Box 439027, San Diego, CA 92143-9027, USA)}
\affil{$^{11}$ Center for Astrophysics and Space Sciences, University of California at San Diego, Code 0424, 9500 Gilman Drive, La Jolla, CA 92093, USA}
\affil{$^{12}$ INAF-Osservatorio Astronomico di Bologna, Via Ranzani 1, I-40127 Bologna, Italy}
\affil{$^{13}$ Max-Planck-Institut f\"{u}r Extraterrestrische Physik, Giessenbachstrasse 1, D-85748 Garching bei M\"{u}nchen, Germany}
\affil{$^{14}$ Harvard Smithsonian Center for Astrophysics, 60 Garden Street, Cambridge, MA 02138, USA}
\affil{$^{15}$ Laboratoire d'Astrophysique de Marseille, BP 8, Traverse du Siphon, 13376 Marseille Cedex 12, France}
\affil{$^{16}$ LBNL \& Berkeley Center for Cosmological Physics, University of California, CA 94720, USA}
\affil{$^{17}$ Department of Physics, ETH Z\"{u}rich, CH-8093 Z\"{u}rich, Switzerland}
\affil{$^{18}$ Max-Planck-Institut f\"{u}r Plasmaphysik, Boltzmanstrasse 2, D-85741 Garching, Germany}
\affil{$^{19}$ Research Center for Space and Cosmic Evolution, Ehime University, Bunkyo-cho, Matsuyama 790-8577, Japan}

\altaffiltext{$\star$}{Based on observations with the NASA/ESA {\em Hubble Space Telescope}, obtained at the Space Telescope Science Institute, which is operated by AURA Inc, under NASA contract NAS 5-26555;
the XMM-Newton, an ESA science mission with instruments and contributions directly funded by ESA Member States and NASA;
European Southern Observatory under Large Program 175.A-0839;
and the Subaru Telescope, which is operated by the National Astronomical Observatory of Japan.
}

\altaffiltext{20}{Member of the IMPRS for Astronomy and Cosmic Physics at the University of Heidelberg}


\begin{abstract}
What is the relevance of major mergers and interactions as triggering mechanisms for active galactic nuclei (AGN) activity?
To answer this long-standing question, we analyze 140 \textit{XMM-Newton}-selected AGN host galaxies and a matched control sample of 1264 inactive galaxies over $z \sim 0.3$--1.0 and $M_\ast<10^{11.7}M_{\odot}$ with high-resolution \textit{HST}/ACS imaging from the COSMOS field.
The visual analysis of their morphologies by 10 independent human classifiers yields a measure of the fraction of distorted morphologies in the AGN and control samples, i.e., quantifying the signature of recent mergers which might potentially be responsible for fueling/triggering the AGN.
We find that (1) the vast majority ($>$85\%) of the AGN host galaxies do not show strong distortions, and (2) there is no significant difference in the distortion fractions between active and inactive galaxies.
Our findings provide the best direct evidence that, since $z\sim 1$, the bulk of black hole (BH) accretion has not been triggered by major galaxy mergers, therefore arguing that the alternative mechanisms, i.e., internal secular processes and minor interactions, are the leading triggers for the episodes of major BH growth.
We also exclude an alternative interpretation of our results: a substantial time lag between merging and the observability of the AGN phase could wash out the most significant merging signatures, explaining the lack of enhancement of strong distortions on the AGN hosts. We show that this alternative scenario is unlikely due to:
(1) recent major mergers being ruled out for the majority of sources due to the high fraction of disk-hosted AGN, (2) the lack of a significant X-ray signal in merging inactive galaxies as a signature of a potential buried AGN, and (3) the low levels of soft X-ray obscuration for AGN hosted by interacting galaxies, in contrast to model predictions.
\end{abstract}



\section{Introduction}
There is a general agreement that supermassive black holes (BHs) lie at the centers of nearly all galaxies, or at least those with a bulge component.
Additionally, strong correlations exist between the BH mass and various properties of the galactic bulge, including luminosity \citep{mbh_l1, mbh_l2}, stellar velocity dispersion \citep{mbh_sigma1, mbh_sigma2, mbh_sigma3}, and stellar mass \citep{mbh_m1, mbh_m2}.
While it has been recently proposed that these correlations are just the product of a statistical convergence of several galaxy mergers over cosmic time \citep{peng07, jahnke&maccio10_astroph},
these correlations have often been interpreted as the signature of coupled evolution between the BH and its host galaxy \citep{kauffmann00,volonteri03,wyithe03,granato04,hopkins07_bhplane,somerville08}.

Given that most galaxies are believed to have undergone a quasar phase, and that the central BH represents a relic of this event \citep{lynden-bell67, richstone98}, the co-evolution picture is naturally very appealing even while some aspects of it remain unclear.
It has been suggested that most of the mass of the BH is built up during the brightest periods of this quasar phase \citep{soltan82, yu&tremaine02}.
If there is such a connection between the growth of the BH and its host galaxy, periods of quasar activity should occur alongside the growth of the bulge, and the mechanism that triggers the accretion onto a once quiescent BH, turning it into an active galactic nucleus (AGN), should be tightly linked with the overall evolution of the host galaxy. The nature of AGN triggering is therefore of key importance for our understanding of galaxy evolution in general.

According to the current paradigm of hierarchical structure formation, major mergers are a crucial element in the assembly and growth of present-day galaxies \citep[e.g.,][]{kauffmann93,cole00,somerville01,bell06,jogee09,robaina10}.
A closer look into the behavior of simulated collisions between galaxies, beginning with the pioneering work of \citet{toomre&toomre72}, suggests that gravitational interactions are an efficient way of transporting material toward the very center of a galaxy.
Mergers and strong interactions can induce substantial gravitational torques on the gas content of a galaxy, depriving it of its angular momentum, leading to inflows and the buildup of huge reservoirs of gas in the center \citep{hernquist89, barnes&hernquist91, barnes&hernquist96, mihos&hernquist96, springel05, cox06, dimatteo07, cox08}.

From early on, major mergers have been related to observations of powerful nuclear starbursts \citep{gunn79}, and connections with quasar activity were made soon after.
\citet{stockton82}, in a study of luminous quasars with close companions, suggested that these neighboring galaxies could be survivors of a strong interaction with the quasar.
Further observational studies came to support this picture:
more cases of quasars with close companions were found, and post-merger features were detected in the host galaxies, whenever it was possible to resolve them \citep[e.g.,][]{heckman84, gehren84, hutchings84, hutchings88, stockton91, hutchings&neff92}.
The merger--quasar connection scenario gained strength with the discovery of the ultraluminous infrared galaxies (ULIRGs).
More than 95\% of these were found in a merging state, some of them hosting an AGN.
This led to the scenario in which ULIRGs and quasars were part of the same chain of events \citep{sanders88a, sanders88b, sanders&mirabel96, surace98, surace99, surace00, canalizo&stockton00, canalizo&stockton01}.

With the advent of the \textit{Hubble Space Telescope} (\textit{HST}), deep imaging of AGN host galaxies at higher redshifts became possible with unprecedented resolution.
Many observational studies of luminous AGN found a high rate of merging signatures in their hosts and detected the presence of very close companions, which before \textit{HST} could not be resolved \citep[e.g.,][]{bahcall97, canalizo&stockton01, zakamska06, urrutia08}.
At the same time, deeper imaging of AGN host galaxies that were initially classified as undisturbed revealed post-merger features not previously detected, both from space-based \citep{canalizo07, bennert08} and ground-based observations \citep{ramosalmeida10}.

There is, however, one major caveat for most of the studies listed above:
almost none of them made use of, or had the access to, an appropriate control sample of inactive galaxies;
such a control sample is essential for discerning if the merger rate is in fact enhanced with respect to the ``background level'', i.e., the merger rate of inactive galaxies.
Only \citet{dunlop03} compared their statistically complete sample of quasars against the quiescent galaxy population, finding no difference in the structural parameters between samples, as well as no enhancement in the large-scale disturbances.
Even if not explicitly, this showed a clear divergence from previous studies regarding the merger-AGN connection scenario, and agreed with the very low frequency of post-merger signatures observed on Seyfert galaxies and low-luminosity AGN \citep{malkan98,schade00}.

A new era of large {\em HST} programs now offers the potential for resolving this discrepancy.
The imaging of larger, contiguous fields has yielded a large number of objects, making it possible to study AGN hosts at space-based resolution, and at the same time to compile a control sample of non-active galaxies.
Initial studies using \textit{HST} imaging by \citet{sanchez04} with the Galaxy Evolution from Morphologies and SEDs survey \citep[GEMS,][]{gems} and by \citet{grogin05} with the Great Observatories Origins Deep Survey \citep[GOODS,][]{goods} found no evidence for an enhancement in merging signatures of AGN hosts over control galaxy samples.
If merger activity does not play a major role in AGN triggering, other methods to produce gas inflows, build up the bulge, and fuel the BH should also be of importance.
Alternate secular mechanisms---minor interactions, large scale bars, nuclear bars, colliding clouds, supernova explosions---can also lead to angular momentum removal and gas inflows from different scales to the central regions \citep[for reviews, see][]{kormendy&kennicutt04,wada04,martini04,jogee06}.
While these processes have usually been related to Seyfert galaxies and low-luminosity AGN \citep[e.g.,][]{simkin80,taniguchi99, hopkins09_fueling}, they could potentially play a larger role than usually reckoned for more luminous AGN as well.
Although the results from the GEMS and GOODS surveys are highly intriguing, the field sizes of $\sim$0.22 deg$^2$ and $\sim$0.08 deg$^2$ respectively were still too small for definitive conclusions to be drawn.
A suitably larger sample would be required to turn these appealing hints into statements.

In this context, we tackle this long-standing issue by performing a comprehensive morphological analysis of a sample of X-ray-selected AGN host galaxies from the Cosmic Evolution Survey \citep[COSMOS,][]{cosmos}, the largest contiguous area ever imaged with the {\em HST} \citep{cosmos_hst, cosmos_acs}.
Our goal is to disentangle the actual relevance and predominance of major galaxy mergers from the other suggested mechanisms for the fueling of the BH.

In the past, targeted high-resolution imaging of AGN hosts has only been possible for small samples, while extensive ground-based surveys with large samples have lacked of the necessary resolution to perform detailed morphological studies at moderate redshifts.
Earlier results from the detailed analysis by \citet{gabor09}, where the morphologies of $\sim$400 AGN host galaxy candidates from the COSMOS field were parameterized, showed that these had an asymmetry distribution consistent with that of a control sample of inactive galaxies, and lacked an excess of companions, already suggesting that major interactions were not predominant among AGN as a triggering mechanism.
Here we use the largest sample of optically confirmed X-ray-selected AGN ever imaged at \textit{HST} resolution from the COSMOS survey and perform a visual inspection of the morphologies of the host galaxies.
We opt for a visual analysis of our galaxies over an automatic classification system because of the inherent problems and incompleteness of the latter in identifying mergers, even for some obvious cases, as cautioned by recent studies probing both methods \citep{jogee09, kartaltepe10b}.
To establish the relevance of our findings, we compare the AGN hosts to a matching sample of inactive galaxies from the same exact data set.

Throughout this paper we assume a flat cosmology with $H_0$ = 70 km s$^{-1}$ Mpc$^{-1}$, $\Omega_{M}$ = 0.3, and $\Omega_{\Lambda}$ = 0.7.
All magnitudes are given in the AB system unless otherwise stated.

\section{Data Set and Sample}

We will perform our analysis on a complete sample of X-ray selected optically confirmed type-1 and type-2 AGN from the COSMOS field.

The COSMOS survey features the largest contiguous area ever imaged with the {\em HST}.
The location of the 1.64 deg$^2$ field, close to the celestial equator, allows access from several major space and ground-based observatories, enabling a large multiwavelength coverage from X-ray to radio from supplementary observational projects.

One of the most effective ways of finding AGN is to make use of the X-ray emission due to the accreting BH \citep[e.g.,][]{mushotzky04}.
Complete coverage of the whole COSMOS field in X-rays was achieved with the \textit{XMM-Newton} \citep[XMM-COSMOS,][]{cosmos_xmm1, cosmos_xmm2} through 55 pointings with a total exposure time of $\sim$1.5 Ms.
We use the catalog presented in \citet{cosmos_brusa10}, which provides the most likely optical and infrared counterparts to the \textit{XMM} sources based on a likelihood ratio technique (see \citealt{cosmos_brusa07} for details).

From the X-ray catalog we draw a parent sample of $\sim$550 sources classified as type-1 AGN from spectroscopic surveys \citep{cosmos_trump07, cosmos_trump09, cosmos_lilly07} revealing broad emission lines, and from spectral energy distribution (SED) fitting \citep{cosmos_capak07, cosmos_salvato09, cosmos_ilbert09}.
We also include a subsample of X-ray selected type-2 AGN, based on those used by \citet{gabor09} drawn from a parent sample of $\sim$300 narrow emission line objects \citep{cosmos_trump07, cosmos_trump09}.

In this paper, we analyze the morphological properties of the AGN host galaxies.
For this, we take advantage of the high-resolution imaging of the COSMOS field with the {\em HST}.
These observations comprise 583 orbits using the Advanced Camera for Surveys (ACS) with the F814W (broad $I$-band) filter \citep{cosmos_acs}.
The imaging data feature an oversampled scale of 0\farcs03/pixel.
Although the ACS survey of the COSMOS field is highly homogeneous, the exact depth achieved is dependent on the angle of the telescope with the Sun at the time of the observations \citep{cosmos_leauthaud07}.
Ninety six out of the 575 pointings were made with an angle smaller than a critical value of $70^{\circ}$, leading to a slightly shallower image.
The limiting surface brightness levels above the background for the pointings made with an angle with the Sun larger and smaller than the critical value are $\sim$23.3 mag arcsec$^{-2}$ and $\sim$22.9 mag arcsec$^{-2}$ respectively.

We restrict our sample to the redshift range $z\sim 0.3 - 1.0$.
For the majority of our final sample, we used high-confidence spectroscopic redshifts, while for the rest (20\%), we used photometric redshifts by \citet{cosmos_salvato09}.
The lower redshift cut is chosen due to the low number of AGN below $z\sim 0.3$ (see Section 3 for further details), and also to avoid working with saturated sources.
The upper limit arises because the F814W filter is shifted into rest-frame UV for sources above $z\sim 1$.
This would mean that we would be specifically looking at the light from young stars and star formation knots, and therefore at biased morphologies.
At the same time, for the case of the type-1 AGN, the bright nucleus would start to dominate due to its blue color, strongly outshining the host and making it nearly impossible to resolve.

\begin{figure}[t]
\centering
\resizebox{\hsize}{!}{\includegraphics{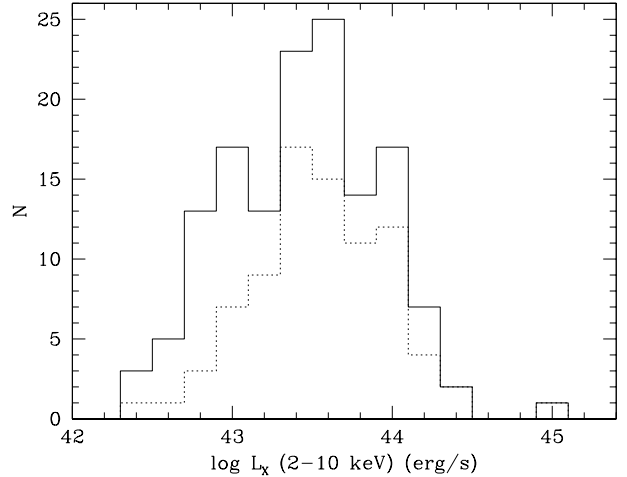}}
\caption{The X-ray luminosity distribution of our sample in the 2-10 keV energy band (solid line). For reference, we also show the distribution of the type-1 AGN subsample only (dotted line).\label{figlx}}
\end{figure}

Given our interest in the morphologies of our sample of AGN host galaxies, we decided not to consider galaxies fainter than $I_{F814W}=24$. Visual morphological classification of these objects would be particularly difficult, and we determined that no consistent information could be extracted at this magnitude.
For the case of the type-1 subsample, we applied this criterion \textit{after} the nucleus removal (see Appendix A for details).

This selection yields 83 type-1 and 57 type-2 AGN.
The median redshifts of the type-1 and type-2 subsamples are 0.80 and 0.67 respectively.
The type-2 subsample has a median apparent $I_{F814W}$ of 20.9, slightly brighter than the type-1 subsample with 21.7 (after nucleus removal, see Appendix A) due to the higher median redshift of the latter.

Figure \ref{figlx} shows the X-ray luminosity distribution of our sources in the 2-10 keV energy band.
The values were obtained mainly from those calculated by \citet{lusso10_lx} and are complemented with those by \citet{mainieri07}.
The median of our distribution lies at $L_X=10^{43.5}$ erg s$^{-1}$, which means that we are probing a reasonably luminous representative AGN sample.
For reference, in Figure \ref{figlx} we also show the X-ray luminosity distribution of the type-1 AGN subsample only, which dominates the overall distribution and has a slightly brighter median $L_X$ ($10^{43.6}$ erg s$^{-1}$) than the type-2 subsample ($10^{43.3}$ erg s$^{-1}$).

\section{Methodology}

In this paper we analyze the morphologies of a sample of AGN host galaxies and of a control sample of inactive galaxies using high-resolution {\em HST}/ACS single $I$-band images.
In the following subsections we explain how we built the comparison sample (hereafter CS), the motivation of choosing a visual inspection over an automatic method, and the classification scheme used.

Analyzing the host galaxies of type-1 AGN is complex, due to the presence of the bright active nucleus in the images, that, depending on the contrast, can outshine the host galaxy to different extents.
We overcome this issue through a two-dimensional decomposition of the AGN and its host galaxy, modeling the bright active nucleus with a point-spread function (PSF) and the host galaxy with a \citet{sersic} profile using GALFIT \citep{galfit02,galfit10}, and then removing the nuclear contribution.
This process is described in detail in Appendix A.

\subsection{Comparison Sample}

The large number of galaxies available from the COSMOS {\em HST} observations provides us with the unique opportunity of building a control sample from the same data set that we draw our AGN from.
For our study, we require the control sample to permit us elaborate a comparison regarding distortion features.
On this respect the most relevant parameter is the signal-to-noise ratio (S/N), hence we construct the comparison sample by selecting inactive galaxies matching each AGN both in apparent $I_{F814W}$ magnitude and photometric redshift.
This is both required and sufficient since (1) the S/N determines the visibility of the merger signatures, and (2) while the stellar masses might differ slightly (factors of $\sim$2 at the same distance and brightness), the mass dependence on the merger rate is not strong, with only a modest increase for higher masses \citep{bundy09}.

Specifically, for each AGN host galaxy we select 10 similar comparison galaxies from the COSMOS ACS catalog \citep{cosmos_leauthaud07}.
Each selected comparison galaxy is required to have an $I_{F814W}$ magnitude within a range of $\bigtriangleup I_{F814W}=$ 0.1, and a photometric redshift within a range of $\bigtriangleup z=$ 0.05.
If not enough galaxies were found, the search ranges were increased by 10\%.
On average, 1.8 iterations were performed to find the required number of inactive galaxies for each AGN host.
For the case of the type-1 AGN subsample, the magnitudes of the host galaxies after the removal of the active nucleus are used for the selection of the control sample.

With the inactive comparison sample in hand, we remove galaxies that are unlikely to be AGN host galaxy counterparts \textit{a priori} via an initial visual inspection.
Such galaxies include: (1) bulgeless disks and irregulars, which would represent a low-mass population, having no corresponding  partners on the AGN sample, and (2) for the type-1 subsample, edge-on disks, which could in principle hold an AGN but this would be heavily obscured and therefore not be a type-1.

Finally, the construction of the control sample for the type-1 AGN requires an additional effort.
The nucleus removal process usually leaves residuals in the center which certainly affect any blind classification, making the type-1 AGN host galaxies readily discernible from the control sample.
To resolve this problem, we mock up our selected inactive galaxies as AGN by adding a star in the center as a fake active nucleus, as we describe in Appendix B.
We then apply the same subtraction procedure as for the original type-1 AGN, attempting to make the two samples indistinguishable.
As we show in Appendix C, any effects on the selection of the comparison sample due to flux variations caused by the nucleus subtraction process can be neglected.

Our final comparison sample consists of 1264 galaxies in total.
The $I_{F814W}$ and redshift distributions of the resulting type-1 and type-2 comparison samples are consistent with being drawn from the same parent distribution as the AGN subsamples, even after the removal of the unlikely counterparts described above.
A Kolmogorov-Smirnov test on each couple of $I_{F814W}$ and redshift distributions confirms with probabilities $>38\%$ that the AGN and control samples are consistent among each other (in general $<5\%$ is used to show that two distributions differ).

\begin{deluxetable*}{lccccccccccc}
\tabletypesize{\scriptsize}
\tablecaption{Results from the Visual Analysis by the 10 Classifiers.\label{tbl-1}}
\tablewidth{1\textwidth}
\tablehead{
Classifier & \colhead{MC} & \colhead{KI} & \colhead{KJ} & \colhead{JK} & \colhead{AK} & \colhead{TL} & \colhead{AR} & \colhead{MS} & \colhead{KS} & \colhead{JT} & \colhead{$\mu$} }
\startdata
$N_{AGN}$ &      140 &      140 &      140 &      140 &       22 &       40 &       57 &      140 &       38 &       98 & ... \\
~$N_{type-1}$ &       83 &       83 &       83 &       83 &       22 &       40 &        0 &       83 &       19 &       41 & ... \\
~$N_{type-2}$ &       57 &       57 &       57 &       57 &        0 &        0 &       57 &       57 &       19 &       57 & ... \\
$N_{CS}$ &     1264 &     1264 &     1264 &     1264 &      177 &      357 &      537 &     1264 &      357 &      903 & - \\
\hline
Hubble type\\
$Bulge_{AGN}$ &    25.7\% &    51.4\% &    31.4\% &    20.0\% &    40.9\% &    55.0\% &    29.8\% &    43.6\% &    26.3\% &    37.8\% &    35.2\% $\pm$
 11.0\% \\
$Disk_{AGN}$ &    74.3\% &    48.6\% &    68.6\% &    80.0\% &    59.1\% &    45.0\% &    70.2\% &    56.4\% &    73.7\% &    62.2\% &    64.8\% $\pm$
11.0\% \\
$Bulge_{CS}$ &    24.6\% &    43.3\% &    29.6\% &    25.2\% &    47.5\% &    54.3\% &    29.4\% &    43.6\% &    17.9\% &    40.5\% &    34.3\% $\pm$
 9.5\% \\
$Disk_{CS}$ &    75.4\% &    56.7\% &    70.4\% &    74.8\% &    52.5\% &    45.7\% &    70.6\% &    56.4\% &    82.1\% &    59.5\% &    65.7\% $\pm$
9.5\% \\
\hline
Distortions\\
Dist--0$_{AGN}$ &    62.9\% &    43.6\% &    48.6\% &    56.4\% &    50.0\% &    47.5\% &    71.9\% &    56.4\% &    47.4\% &    55.1\% &    54.2\% $\pm$
     7.5\% \\
Dist--0$_{CS}$ &    65.5\% &    47.3\% &    60.1\% &    63.0\% &    67.8\% &    51.5\% &    78.0\% &    58.5\% &    51.5\% &    61.2\% &    59.9\% $\pm$
    7.6\% \\
$\Delta_{Dist-0}$ &    --2.6\% &    --3.7\% &   --11.6\% &    --6.5\% &   --17.8\% &    --4.0\% &    --6.1\% &    --2.1\% &    --4.2\% &    --6.1\% &    --5.6\% $\pm$     3.5\% \\
Dist--1$_{AGN}$ &    24.3\% &    26.4\% &    45.0\% &    32.9\% &    40.9\% &    40.0\% &    21.1\% &    30.0\% &    50.0\% &    16.3\% &    30.8\% $\pm$
     9.3\% \\
Dist--1$_{CS}$ &    22.4\% &    28.3\% &    34.2\% &    26.9\% &    19.2\% &    34.2\% &    16.0\% &    33.5\% &    39.2\% &    17.8\% &    27.5\% $\pm$
    6.5\% \\
$\Delta_{Dist-1}$ &     1.9\% &    --1.9\% &    10.8\% &     6.0\% &    21.7\% &     5.8\% &     5.0\% &    --3.5\% &    10.8\% &    --1.5\% &     3.2\% $\pm$     5.7\% \\
Dist--2$_{AGN}$ &    12.9\% &    30.0\% &     6.4\% &    10.7\% &     9.1\% &    12.5\% &     7.0\% &    13.6\% &     2.6\% &    28.6\% &    15.0\% $\pm$
     8.8\% \\
Dist--2$_{CS}$ &    12.1\% &    24.4\% &     5.7\% &    10.1\% &    13.0\% &    14.3\% &     6.0\% &     7.9\% &     9.2\% &    20.9\% &    12.6\% $\pm$
    6.5\% \\
$\Delta_{Dist-2}$ &     0.8\% &     5.6\% &     0.7\% &     0.6\% &    --3.9\% &    --1.8\% &     1.1\% &     5.7\% &    --6.6\% &     7.6\% &     2.4\% $\pm$     3.5\%
\enddata
\tablecomments{We indicate the number of objects classified by each person for the AGN sample and individually for each type-1 and type-2 subsample, as well as for the comparison sample (CS).
For the distortion classifications, we include the difference between samples as $\Delta_{Dist-X}$=Dist--X$_{AGN}$ -- Dist--X$_{CS}$.
For each category, we include the mean, $\mu$, and its dispersion, weighted according to the number of objects classified by each person.}
\end{deluxetable*}

\begin{deluxetable}{lccc}
\tabletypesize{\scriptsize}
\tablecaption{Comparison of our mean Hubble type classification with that of parametric estimators of galaxy morphologies\label{tbl-2}}
\tablewidth{0.4\textwidth}
\tablehead{
\colhead{} & \colhead{$\mu$\tablenotemark{a}} & \colhead{GALFIT\tablenotemark{b}} & \colhead{ZEST\tablenotemark{c}} }
\startdata
Bulge$_{AGN}$  &  35.2\% & 25.7\% & ...  \\
Disk$_{AGN}$   &  64.8\% & 55.0\% & ...  \\
Bulge$_{CS}$ &  34.3\% & 41.2\% & 19.6\% \\
Disk$_{CS}$  &  65.7\% & 43.5\% & 67.8\%
\enddata

\tablenotetext{a}{Weighted mean of the 10 classifications (as in Table~\ref{tbl-1}).}
\tablenotetext{b}{Percentages over 100\% of the samples. The rest of the objects had an intermediate S\'{e}rsic index (with $2\leq n\leq3$, see Appendix A for details).}
\tablenotetext{c}{Percentages over 100\% of the comparison sample. The remaining galaxies were classified as irregulars (11.3\%) by ZEST, and a few did not make it to the catalog (1.2\%).}
\end{deluxetable}

\subsection{Visual Classification}

Merger events come in many different flavors due to the large parameter space involved (e.g., merger stage, viewing angles, mass ratio, and gas fractions).
Sometimes they can be obvious at first sight, but some others can be very subtle, or simply undetectable at the sensitivity of the observations.
At our redshift range and image resolution, it has been shown that automatic classification methods to identify mergers tend to miss several obvious cases, and cannot compete with visual inspection \citep{jogee09, kartaltepe10b}.
On the other hand, when the numbers involved are over the tens of thousands, visual classification becomes impractical\footnote{With the notable exception of the citizen-based Galaxy Zoo project \citep[][http://www.galaxyzoo.org]{galaxyzoo}.} and an automatic approach would be needed.
General measurements of structural parameters that can be correlated with some physical process have proven to be a good compromise
\citep[e.g.,][using the lopsidedness as a tracer of merging and star formation]{reichard09}.

Considering the above, in this paper we opt to identify merger and interaction signatures visually.
The number of objects we are dealing with allows us to do so ($\sim$1400 in total), and the image quality deserves a detailed case-by-case examination.

These visual studies can be subjective.
In our case, the absolute fraction of merging galaxies measured by visual classifiers will depend on their own experience and background, and hence it is plausible that they can differ substantially among each other.
Nevertheless, any personal scale and criteria each classifier uses will be applied equally on both samples, active and inactive galaxies.
Therefore, a key quantity on our study will be, more than the absolute fractions of merging galaxies, the difference \textit{between} the merging fractions measured by a given classifier.
If we instead focus on how each individual classifier perceives one sample \textit{compared to} the other, by considering the differential between the merging fractions of active and inactive galaxies, this subjectiveness can be accounted for.
Furthermore, the consistency of this study is improved by
(1) using ten independent human classifiers to add statistical robustness and
(2) mixing both samples of active and inactive galaxies so that the classification is actually blind and therefore does not favor either the AGN hosts or the inactive galaxies.

\begin{figure*}[t]
\centering
\begin{tabular}{l|cc}
&AGN host galaxies&Inactive galaxies\\ \hline
&&\\
\begin{sideways}Bulge-dom.\end{sideways}&
\begin{tabular}{cc}
 \includegraphics[width=0.15\textwidth]{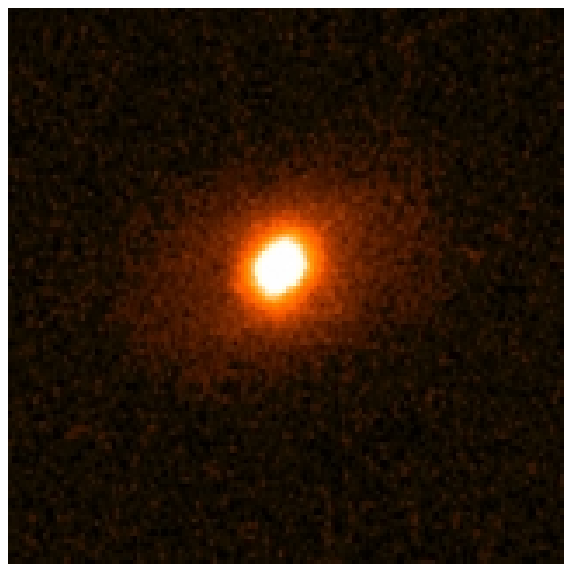}&
 \includegraphics[width=0.15\textwidth]{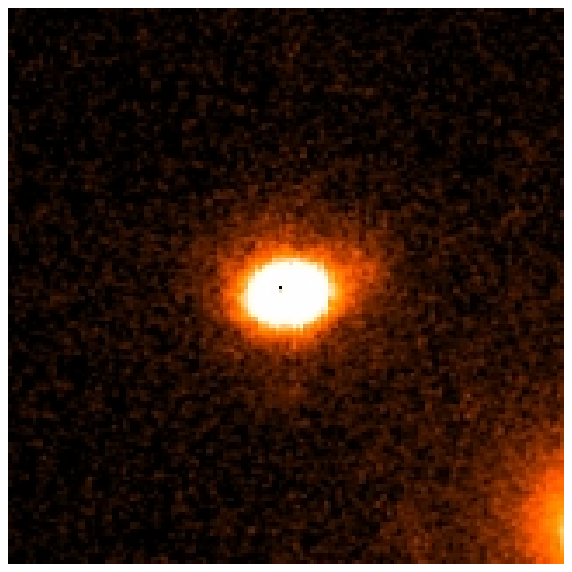}\\
\end{tabular}&
\begin{tabular}{cc}
 \includegraphics[width=0.15\textwidth]{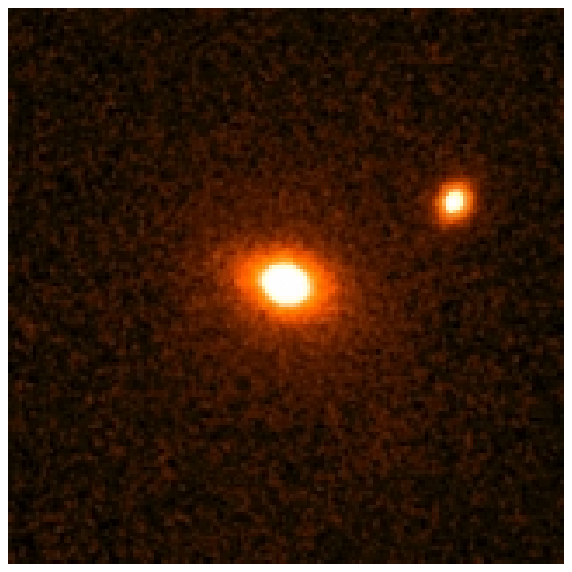}&
 \includegraphics[width=0.15\textwidth]{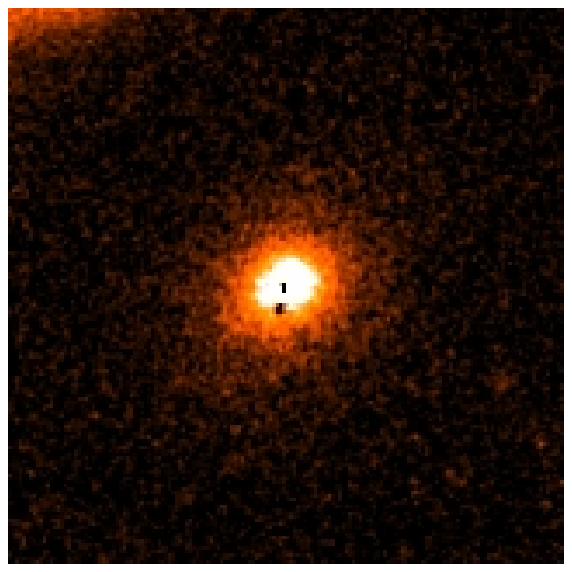}\\
\end{tabular}\\

\begin{sideways}Disk-dom.\end{sideways}&
\begin{tabular}{cc}
 \includegraphics[width=0.15\textwidth]{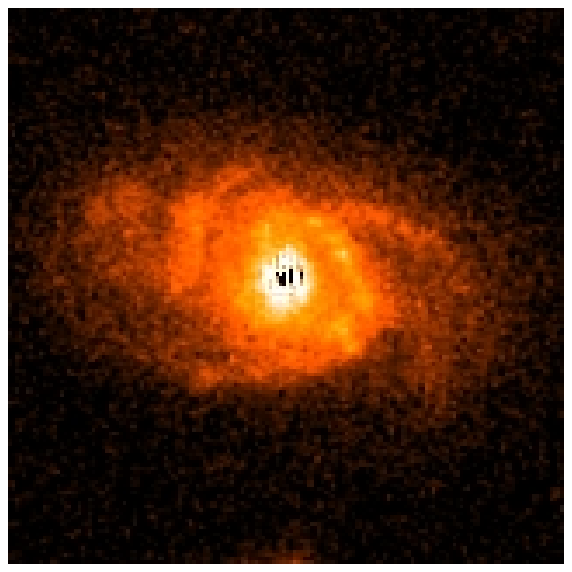}&
 \includegraphics[width=0.15\textwidth]{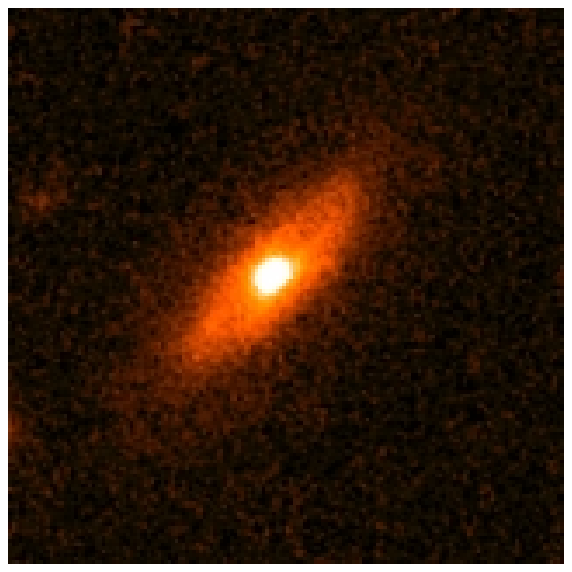}\\
\end{tabular}&
\begin{tabular}{cc}
 \includegraphics[width=0.15\textwidth]{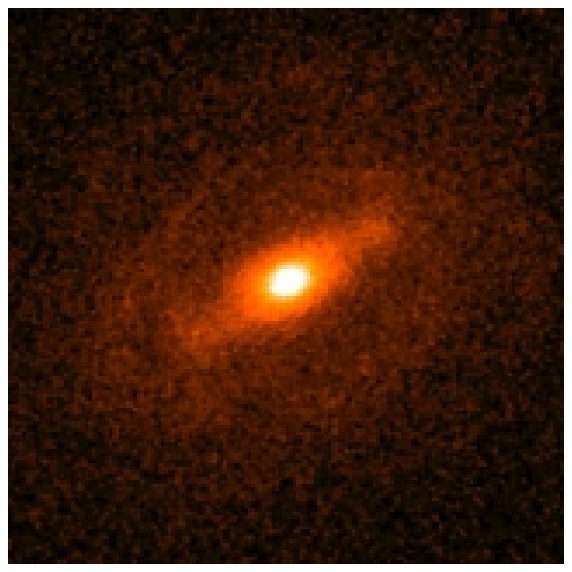}&
 \includegraphics[width=0.15\textwidth]{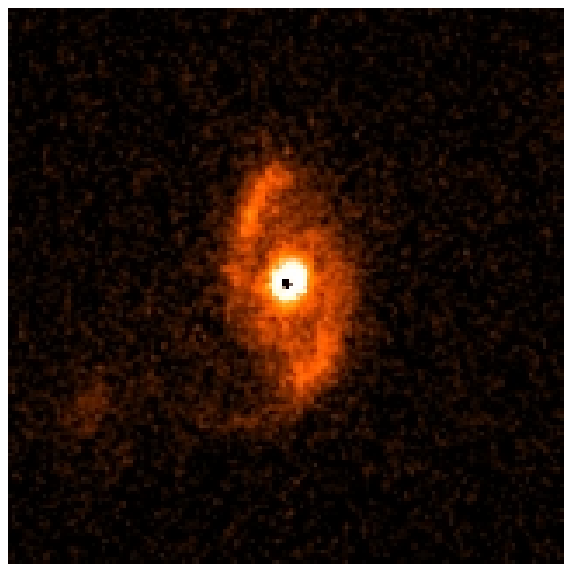}\\
\end{tabular}\\

\end{tabular}

\caption{\label{fig_hubble}
Example galaxy images arranged into different morphological classes with 100\% agreement between the independent classifiers. The cutouts are $4\farcs8$ $\times$ $4\farcs8$. Black residuals at the center of some of the galaxies are residuals from the point source removal.}
\end{figure*}

\begin{figure*}[t]
\centering
\begin{tabular}{l|cc}
&AGN host galaxies&Inactive galaxies\\ \hline
&&\\
\begin{sideways}Dist-0\end{sideways}&
\begin{tabular}{cc}
 \includegraphics[width=0.15\textwidth]{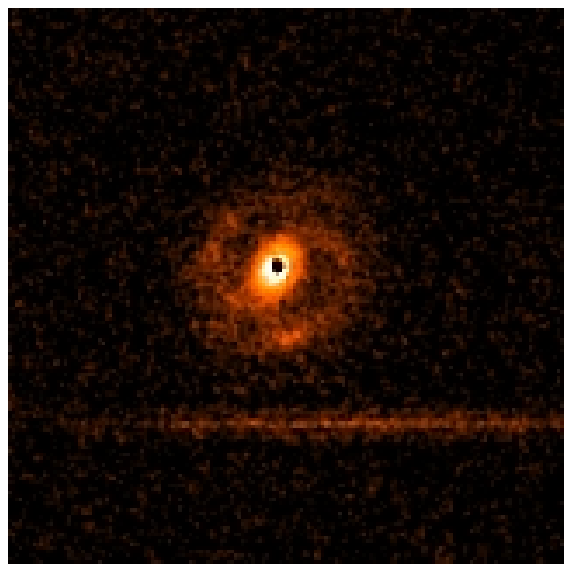}&
 \includegraphics[width=0.15\textwidth]{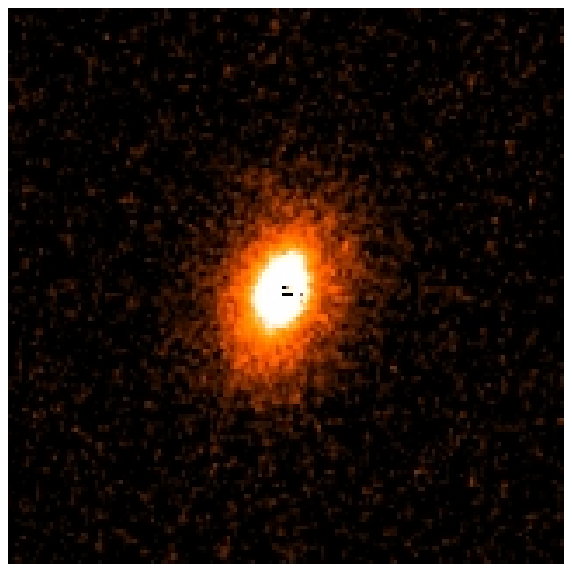}\\
\end{tabular}&
\begin{tabular}{cc}
 \includegraphics[width=0.15\textwidth]{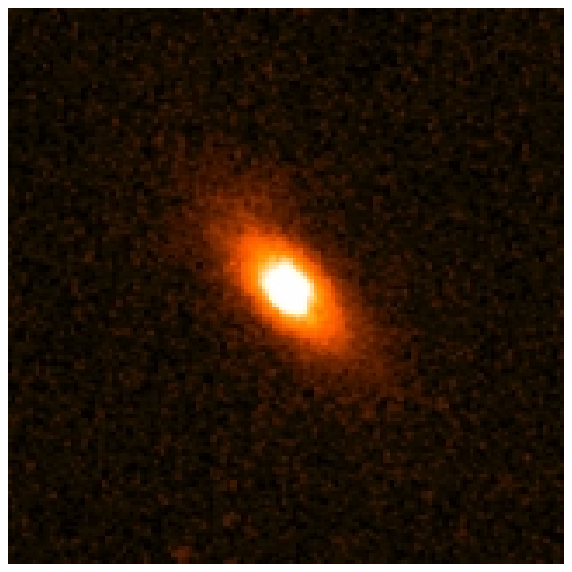}&
 \includegraphics[width=0.15\textwidth]{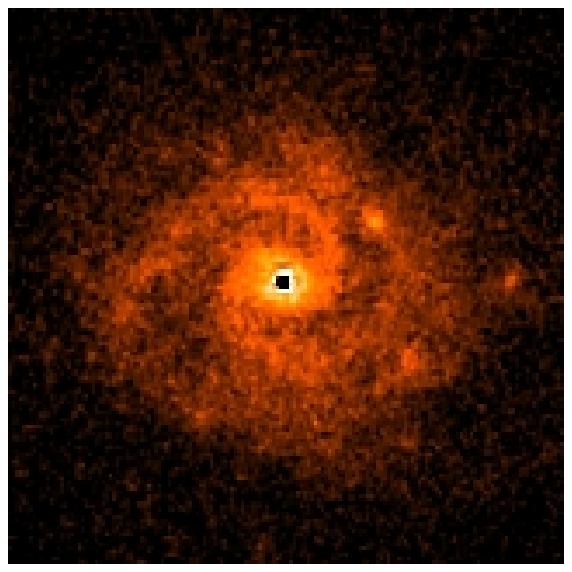}\\
\end{tabular}\\

\begin{sideways}Dist-1\end{sideways}&
\begin{tabular}{cc}
 \includegraphics[width=0.15\textwidth]{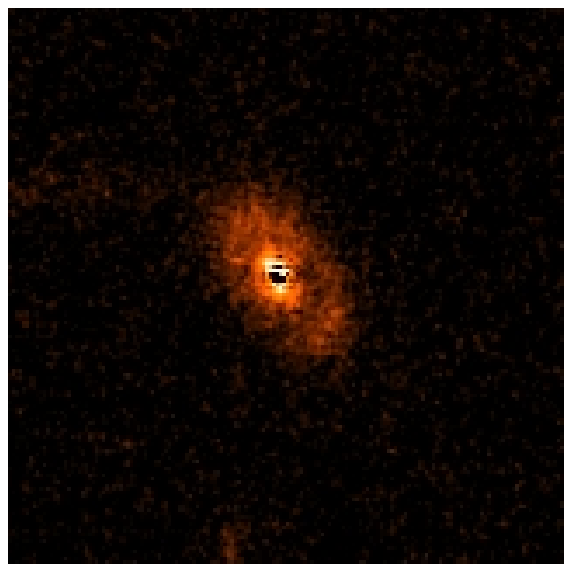}&
 \includegraphics[width=0.15\textwidth]{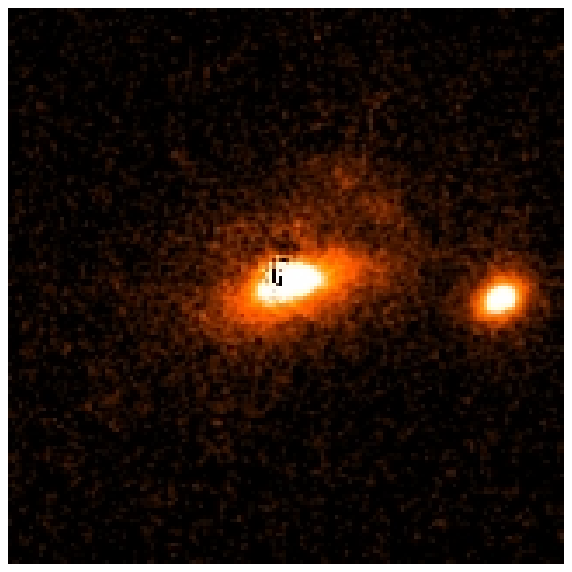}\\
\end{tabular}&
\begin{tabular}{cc}
 \includegraphics[width=0.15\textwidth]{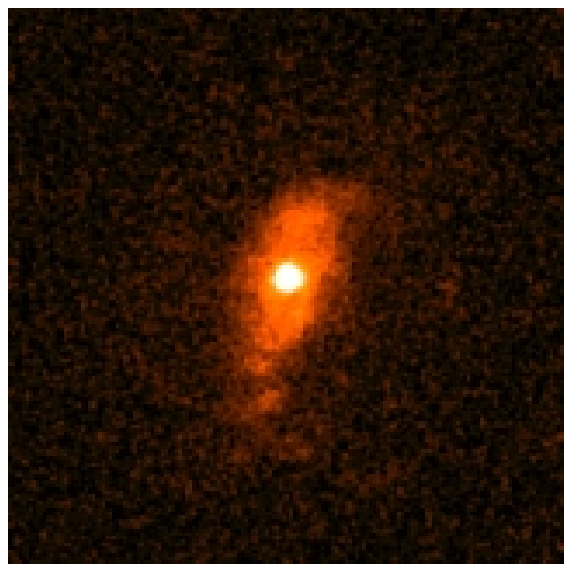}&
 \includegraphics[width=0.15\textwidth]{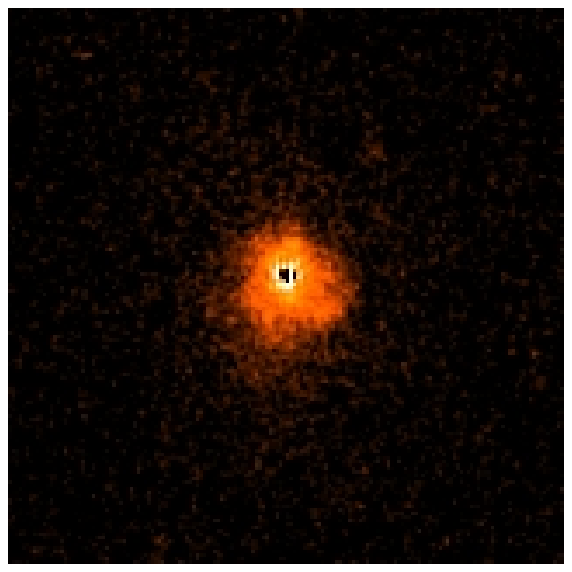}\\
\end{tabular}\\

\begin{sideways}Dist-2\end{sideways}&
\begin{tabular}{cc}
 \includegraphics[width=0.15\textwidth]{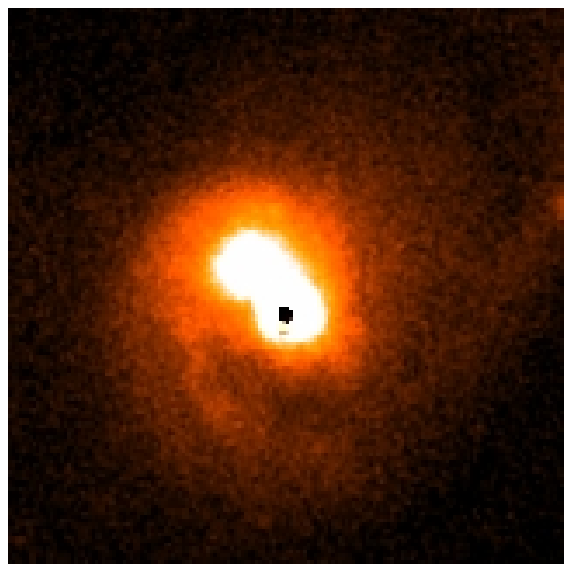}&
 \includegraphics[width=0.15\textwidth]{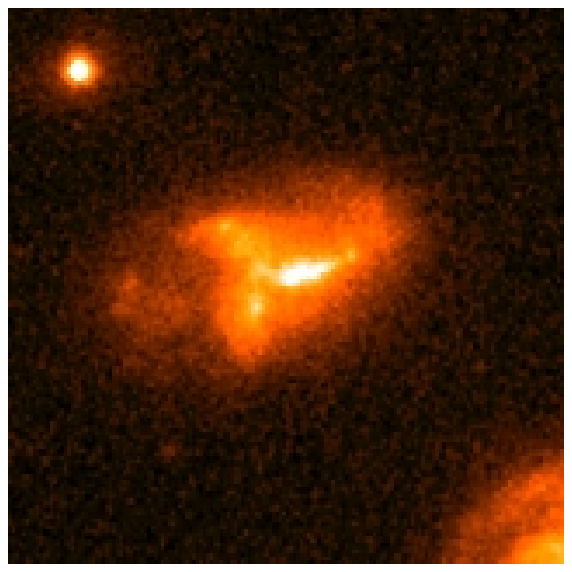}\\
\end{tabular}&
\begin{tabular}{cc}
 \includegraphics[width=0.15\textwidth]{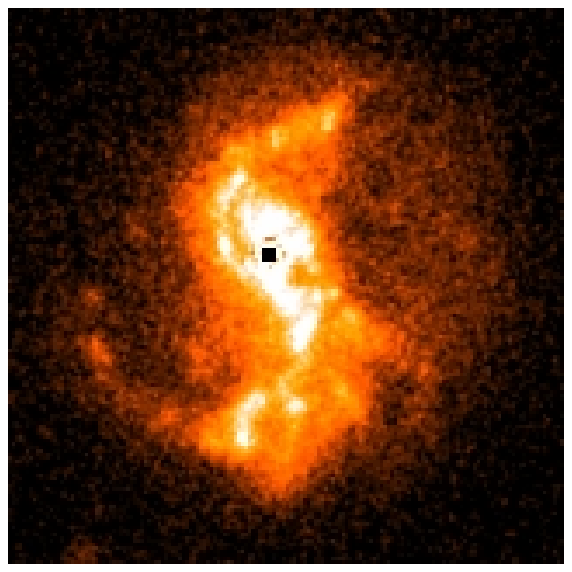}&
 \includegraphics[width=0.15\textwidth]{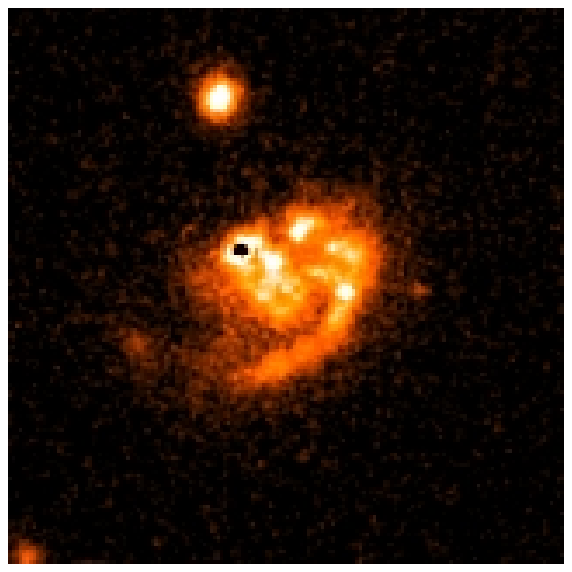}\\
\end{tabular}\\

\end{tabular}

\caption{\label{fig_dist}
Example galaxy images arranged into different distortion classes with 100\% agreement between the independent classifiers. The cutouts are $4\farcs8$ $\times$ $4\farcs8$. Black residuals at the center of some of the galaxies are residuals from the point source removal.}
\end{figure*}

We break the classification down into two parameters.
\begin{itemize}
 \item[1.] \textit{Hubble type.} We attempt to state whether the host galaxy belongs to one of the two basic morphological classes: bulge or disk dominated.
 \item[2.] \textit{Distortion class.} We define three classes regarding the degree of distortion of the galaxy as follows.
 \begin{itemize}
  \item[(a)] \textit{Dist-0.} Galaxies that appear undisturbed, smooth and/or symmetric, showing no interaction signatures. This also applies to cases where the small diameter of the galaxy does not allow a detailed analysis. We pay particular attention to self-induced asymmetries such as dust lanes or star-forming regions, which are usually seen as small clumps in well-resolved spirals.
  \item[(b)] \textit{Dist-1.} Here we include objects with mild distortions. This could be due to a minor merger for example, but at the same time could also be because of low S/N. This interaction class is a "gray zone" in which most of the discrepancies in the classification between the 10 people arise.
  \item[(c)] \textit{Dist-2.} Strong distortions, potential signs for ongoing or recent mergers. This class mainly includes galaxies which have highly disturbed morphologies or show visible signatures of strong interactions, such as large tidal tails, arcs, debris, etc. Double-nucleus systems also fall into this category.
 \end{itemize}
\end{itemize}

Illustrative examples of the Hubble type and distortion classes are shown in Figures \ref{fig_hubble} and \ref{fig_dist}, respectively.

For the visual inspection, the classifiers had access to FITS images which they could re-scale in order to look for high-contrast and subtle features that may have not showed up at an arbitrary brightness scale.

\section{Results}

The results from the visual classification by 10 people (MC, KI, KJ, JK, AK, TL, AR, MS, KS, and JT), for both Hubble type and distortion classes, are shown in Table~\ref{tbl-1}.
For the different distortion classes, we show the difference between samples (hereafter $\Delta$) as the distortion fraction of the AGN minus that of the control sample.
The results are weighted according to the number of objects classified by each person\footnote{Each classifier looked at a minimum of $\sim$200 galaxies from the combined sample. For each classifier the samples were shuffled, to assure that even if all of them decided to look at 200 galaxies, they would be looking at different objects. On average, each galaxy was classified 6.3 $\pm$ 1.0 times.} and used to calculate the mean fractions, $\mu$, which we also display in Table~\ref{tbl-1}.
Figures \ref{fig_hubble} and \ref{fig_dist} show examples of active and inactive galaxies which were classified with 100\% agreement, arranged into the different Hubble type and distortion classes respectively.

\begin{figure*}[t]
\centering
\resizebox{0.85\textwidth}{!}{\includegraphics{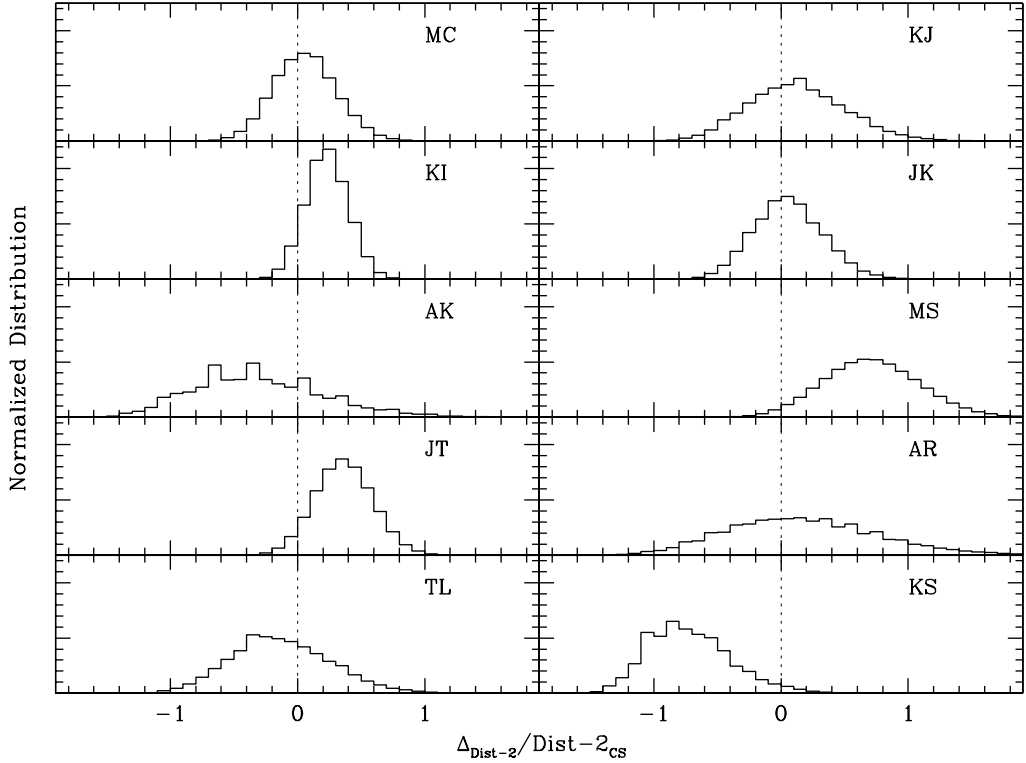}}
\caption{Distributions of the difference in the Monte Carlo sampled distributions of Dist-2 fractions between the AGN and control samples for the ten classifiers. For each distribution, a deviation from zero difference (dotted line) towards positive values indicates a higher fraction of distorted active galaxies, whereas a deviation towards negative values shows a higher fraction of distorted inactive galaxies.\label{fig_diffind}}
\end{figure*}

\subsection{Perception of the Hubble type}

No morphology priors are applied in the selection of our comparison sample, with the minor exception  of the pruning of irregulars and edge-on disks as described in Section 3.1.
In order to test whether the samples are consistent regarding their morphological composition, we compare the AGN and comparison samples in Table~\ref{tbl-1}.
Although the mean values show a high dispersion due to the large discrepancies between classifiers, the results for the AGN and comparison samples are in good general agreement for each classifier.

The large fraction of disks, in particular in the AGN sample, is interesting.
To verify if this could be due to systematic bias by the classifiers, we will use two independent parametric estimators of the morphological type available at hand.
First we compare the results from our GALFIT models chosen earlier, which we extended to our type-2 subsample as well as to its comparison galaxies.
We identify sources as bulge- or disk-dominated if the best-fit results from GALFIT had S\'{e}rsic indices of $n=4$ and $n=1$ respectively. The rest of the galaxies fell between the two.
As a second test, we look up the results for our comparison sample from the Zurich Estimator of Structural Types (ZEST, see \citealt{zest} for details), in which the structure of thousands of COSMOS galaxies was quantified through a principal component analysis over a combination of S\'{e}rsic index and five non-parametric diagnostics.
The ZEST results show the fractions of galaxies classified either as bulges or disks. Of the remaining fraction classified as neither, the majority (11.3\%) was classified as irregulars, most likely due to the lack of sensitivity of these automatic classification schemes to peculiar systems such as interacting galaxies; this is consistent with the observed fraction of highly distorted comparison galaxies.
Sixteen galaxies, accounting for the remaining 1.2\%, did not make it into the catalog.

Table~\ref{tbl-2} shows both of these tests along with the weighted mean fractions for comparison.  It is clear that the numbers from these tests follow the trend seen in the visual classification.
These tests provide a lower limit to the fraction of disks, with $>55\%$ of our AGN sample being hosted by true disks.

\subsection{The distortion fractions}

Our prime interest lies in the observed difference in distortion fractions \textit{between} samples of active and inactive galaxies.
The absolute values in distortion fractions determined by the 10 classifiers are of lesser interest since the internal calibration for the three distortion classes differs between the classifying individuals.
Since any subjectiveness will be applied equally to both active and inactive samples, using the differences in the distortion fractions instead of absolute levels removes the person-to-person calibration differences and allows an unbiased interpretation.

Considering that the merging signatures we were looking for could sometimes be faint and weak, we address the potential loss of sensitivity to such features due to the slightly shallower limiting magnitudes for $\sim$17\% of the pointings (i.e., those with Sun-angles of $<$70$^\circ$).
For each person, we have also carefully analyzed the results by dividing their classified sample into sources with sun-angles either side of this critical angle.
We find that there is no statistically significant difference in the distortion fractions as a function of Sun-angle.
In addition, as the assignment of individual objects to either a deep or shallow field is effectively random, and given that the AGN distortion fractions are compared directly with those of a comparison sample selected from the same data set (and thus with the same limiting surface brightness issues), the overall impact on our results of any bias toward smaller distortion fractions in the shallower fields would in any case be negligible.

The objects that fell into the Dist-2 class were those which presented the strongest distortions, and hence signatures of a major interaction, for each individual classifier
As any difference in recent major merger incidence would show itself in this class, we will focus on the Dist-2 results here.

\begin{figure}[t]
\centering
\resizebox{\hsize}{!}{\includegraphics{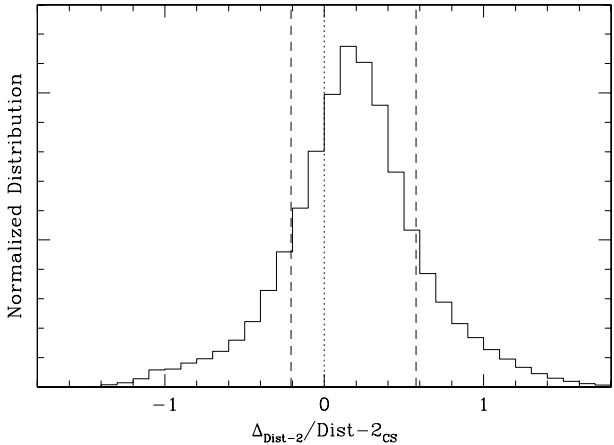}}
\caption{Combined posterior probability distribution of the difference of highly distorted galaxies between the AGN and control sample for the 10 classifiers. The central 68\% confidence level is marked with vertical dashed lines, which shows that the histogram is consistent with zero difference (dotted line), ruling out any significant enhancement of merging signatures on our sample of AGN hosts with respect to the comparison sample of inactive galaxies.\label{fig_diffall}}
\end{figure}

\begin{figure}[t]
\centering
\resizebox{\hsize}{!}{\includegraphics{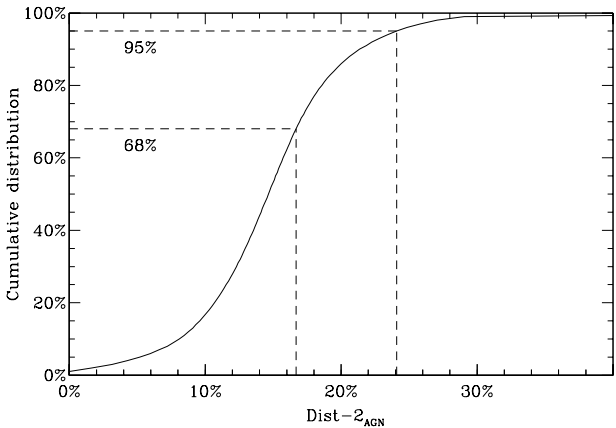}}
\caption{Cumulative distribution of the simulated Dist-2$_{AGN}$ fractions, showing the 68\% and 95\% confidence levels with the dashed lines. As mentioned in the text, this confirms with a 95\% confidence that the highly distorted AGN fraction can \textit{not} be larger than 24.08\%.\label{fig_confall}}
\end{figure}

\subsubsection{Combining 10 classifications}

In table~\ref{tbl-1}  we have already listed the Dist-2 fractions for all classifiers, their mean values, and also the mean of the difference in Dist-2 fractions between the AGN and comparison samples.
This permits the following initial assessment under the assumption of Gaussian errors:
the difference (2.4\%) is below the uncertainty of 3.5\%, and hence it is not significant.

Nevertheless, since the error distribution is in fact \textit{not} Gaussian but follows a \textit{binomial} distribution (according to the number of distorted galaxies in a sample of given size) it is important that we use the correct combination of results in order to give answers to the two main questions:
(1) Is there a genuine difference between the fractions of strongly distorted AGN hosts and inactive galaxies, and
(2) with the given sample size, what difference in distortion fractions between samples can we actually rule out at a given confidence level---in this case we chose 95\%.
The first question asks whether the given dataset shows an enhanced AGN distortion fraction or not.
The second question probes the discriminative power of this sample, and allows us to gauge the actual importance of a null-result in question (1), since a decreasing sample size means an increasing uncertainty in the distortion fractions and hence small samples have near zero discriminative power.

Using the correct binomial error statistics for the distortion fractions of AGN and inactive galaxies, we compute for each classifier the probability distribution for the difference $\Delta_{Dist-2}$.
This is done in the following way:
(1) individially for each classifier, we Monte Carlo sample their pair of Dist-2 binomial probability distributions for the AGN and comparison samples separately,
(2) we compute the difference between these randomly sampled values,
(3) we repeat this process one million times for each classifier which yields 10 distributions for $\Delta_{Dist-2}$,
(4) we normalize these probability distributions by the Dist-2$_{CS}$ values measured by each person as shown in Table~\ref{tbl-1}, in this way removing any bias applied by each classifier's personal scale (Figure~\ref{fig_diffind}),
and (5) now in ``differential'' space, where we are insensitive to between-person scatter, we combine these 10 distributions by co-adding their histograms, weighted by the size of the sample each person classified\footnote{This represents a combined Bayesian posterior probability distribution with sample sizes as individual priors.}.

The resulting probability distribution is shown in Figure~\ref{fig_diffall}.
The histogram is fully consistent with zero difference, as indicated by the central 68\% confidence interval denoted by the vertical dashed lines estimated by the areas at both ends, encompassing 16\% each.
This confirms the simple analysis from above:
our study shows no significant difference between the fractions of strong distortions of AGN and inactive galaxies.
Regarding the discriminative power of our sample, in Figure~\ref{fig_confall} we show the cumulative distribution of the Dist-2$_{AGN}$ fraction from Figure~\ref{fig_diffall}.
The distribution shows that with 95\% confidence the distortion fraction of AGN is in any case not larger than the inactive distortion fraction by a factor of 1.9, when considered relative to the mean distortion level found by the 10 classifiers (12.6\%).
Hence, the vast majority of AGN host galaxies at $z<1$ with the given luminosities do not show signatures of having experienced a recent major merger.

\begin{figure}[t]
\centering
\resizebox{\hsize}{!}{\includegraphics{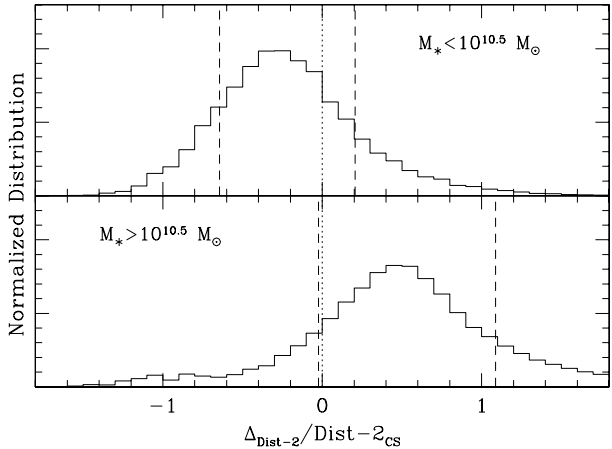}}
\caption{The combined differences in distortions of intermediate ($10^{9.3}<M_{\ast}/M_{\odot}<10^{10.5}$; top panel) and massive ($10^{10.5}\leq M_{\ast}/M_{\odot}<10^{11.7}$; bottom panel) galaxies are shown. In both cases, the central 68\% confidence levels (dashed lines) are consistent with zero (dotted line).\label{fig_diffmass}}
\end{figure}

\subsubsection{Mass dependency}
Even if there is no overall difference between the fractions of highly distorted AGN and inactive galaxies, it is still interesting to look at the situation in mass-space, and investigate the possibility that an enhancement of the AGN merger fraction could be \textit{hidden} because we consider the sample as a whole, regardless of stellar mass.
Major merging is a key element in the assembly and evolution of massive galaxies \citep[e.g.,][]{bell06,lin08,bundy09,vanderwel09,robaina10}, and in order to test if the fraction of highly distorted AGN host galaxies is significantly enhanced at the massive end (higher than $\sim10^{10.5}$ $M_{\odot}$),
we have estimated stellar masses for our samples of active and inactive galaxies.
We use the calibration from \citet{bdj01} based on the Chabrier initial mass function.
By obtaining the $V$-band luminosities:
\begin{equation}
 L_V = 10^{-0.4(V-4.82)}
\end{equation} 
and assuming a common mass-to-light ratio from the rest-frame ($B - V$) color, we derive stellar masses in solar units:
\begin{equation}
 M_\ast = 10^{-0.728+1.305(B-V)} \times L_V
\end{equation}
with all magnitudes in Vega zero point.

For the inactive galaxies and the type-2 subsample we obtain rest-frame $B$ and $V$ from the photometric catalog by \citet{cosmos_ilbert10}.
For the type-1 AGN, however, we cannot use that information because it includes the contribution from the luminous AGN.
Therefore, we obtain the rest-frame $V$-band luminosities from the observed $I_{F814W}$ after the nucleus removal process and estimate the color term by computing the linear regression over the rest-frame ($B - V$) colors as a function of redshift for the type-2 AGN.
This yields the relation
\begin{equation}
 (B - V)_{Vega} = 0.136~ z + 0.541.
\end{equation}

The combined differences of highly distorted galaxies for two bins of stellar mass ($10^{9.3}--10^{10.5} M_{\odot}$ and $10^{10.5}--10^{11.7} M_{\odot}$) are shown in Figure~\ref{fig_diffmass}.
For both samples the ratio of galaxies occupying the massive bin is roughly 2:1 relative to the less massive one, hence we are dealing with very massive galaxies.
Even if for the galaxies with stellar masses higher than $\sim10^{10.5}$ $M_{\odot}$ there is a modest enhancement in the distortion fraction of the AGN hosts over the control sample (Figure~\ref{fig_diffmass}, bottom panel), it is again within the 68\% confidence interval, i.e., it is not statistically significant.
Therefore, it cannot be considered as an empirical proof of an enhancement at the massive end.

\section{Discussion}
From a detailed analysis of the results of our visual classification we showed that the fractions of heavily distorted active and inactive galaxies are consistent within the central 68\% confidence interval and that the Dist-2 fraction of AGN host galaxies is less than twice that of the inactive galaxies at a 95\% significance level, as shown in Figures \ref{fig_diffall} and \ref{fig_confall}, respectively.
Putting these findings in context, provided that the duration of merger signatures and the visibility of the AGN phase overlap with each other, this indicates that there is no evidence that major merging plays a key role in the triggering of AGN activity in our sample.
But what about the possible alternative scenario in which, in spite of a causal connection between merging and AGN triggering, we do not detect an enhancement of merger signatures in the AGN population due to a significant time lag between the interaction and the start of the AGN phase?
Below we address this possible alternative interpretation with some simple tests, and discuss the implications of our results.

\subsection{Alternative Interpretation: Time Lag Between Merging and the Observability of the AGN Phase}

Appealing simulations of mergers between gas-rich galaxies state that the peak of star formation and quasar activity will occur during the final stages of the interaction, close to coalescence, within a more relaxed than distorted bulge-like remnant \citep{dimatteo05, springel05}.
In these models, during the first passage only modest starbursts are triggered and no major BH accretion occurs, and therefore the galaxies would not be detected as AGN.
Furthermore, ad hoc models that include obscuration in galaxy mergers \citep{hopkins05_lifeqso} predict that, beginning from the early stages of the interaction, the AGN is ``buried'' for $\sim$90\% of its lifetime by large column densities, only revealing itself toward the end of the merger.
However, all these models work with sub-grid prescriptions of BH accretion and fail to spatially resolve the actual accretion process by several orders of magnitude.

If there is indeed a substantial time lag after merging prior to the AGN activity becoming detectable, then the strong merging signatures we attempt to find could have already been washed out.
Moreover, if AGN are obscured as the interacting galaxies coalesce, there could be a ``contamination'' population of undetected strong BH activity occurring within our control galaxies undergoing a major merger.
Finally, a third issue related to the obscuration plus time lag scenario is that the observed interactions that are occurring on a fraction ($\sim$15\%) of our AGN host galaxies should be unrelated to the detected BH accretion---under the assumption of a large time lag we would not expect to see strong merging signatures.

\subsubsection{AGN Hosted by Disks: Not a Relic from a Major Merger}

In the preceding text, we raised a possible alternative explanation for our results, that most major mergers could be missed because the time lag between merging and the observed AGN episode could be substantial, washing out the signatures that the \textit{HST}/ACS resolution allows us to detect.

Models can provide us with some clues about the observability timescales during an interaction.
For example, simulations of major mergers by \citet{lotz08} quantified that the strong signatures could still be detected 0.7 Gyr after the merger, by degrading their snapshots to the resolution of \textit{HST} $z\sim 1$ imaging.
Thus, in order to explain the observed zero distortion enhancement, a lag of at least 0.7 Gyr between coalescence and the visible phase of the AGN would be required for all galaxies\footnote{For example, see \citet{schawinski10} who make an extensive case of the time lag scenario. They propose an all-merger-driven AGN phase with a time lag of $\sim$500 Myr for their sample of early-type galaxies at $z\sim0.05$.
Even if their result is mainly based on the interpretation of their data as a causal sequence of events (and is subject to alternative explanations), they caution that their particular sample only accounts for a very small fraction ($\sim$10\%) of the overall AGN population found in the local universe.}.
It is, however, not straightforward to rely on these studies to discard the time lag issue;
given the large number of parameters involved in determining how long a merger signature will remain visible, it is plausible that several late-stage mergers could have been missed.
Although a merger between gas-rich galaxies can leave spectacular features for a long time, viewed from the wrong orientation they can be completely unnoticeable.

While it is difficult to assess the relevance for the timescale issue of major mergers being overlooked, we can be reasonably confident that the remnant will not look like a disk.
Spheroidal and bulge-dominated galaxies are usually said to be formed as a result of major mergers \citep[e.g.,][]{toomre77, barnes&hernquist96, cox06}.
However, it has also been stated that disks can survive some major mergers, especially if the progenitors are gas-rich \citep[e.g.,][]{barnes&hernquist96, springel&hernquist05, hopkins09_survive}, nonetheless these kind of merger remnants have been argued to not lead to a large bulge growth and significant BH fueling \citep{hopkins09_fueling}.
Likewise, it has been argued that some gas-rich mergers can lead to the regrowth of the disk \citep{hopkins09_survive, bundy10}.
Even so, the timescales involved for such a process can be as much as an order of magnitude larger than the typical quasar lifetime of 1-100 Myr \citep[e.g.,][]{porciani04, hopkins05_lifeqso, shen07}.

For the significant fraction of AGN hosted by disks found from our classification, we could safely say that the mechanism responsible for triggering those AGN was not a past major merger, suggesting also that since $z\sim 1$ alternative fueling methods seem to play a larger role than usually expected.
\citet{georgakakis09}, from a sample of X-ray-selected AGN, compared the luminosity function of their disk-hosted AGN against the analytic model of the X-ray AGN luminosity function for a stochastic accretion mode by \citet{hopkins06_stochastic}.
They showed that the model can reproduce the observations, but at the same time the overall number density of the observed disks was underpredicted, especially at high X-ray luminosities.
On our sample of 140 AGN, 18 sources have $L_X\geq10^{44}$ erg s$^{-1}$, from which 10 of their host galaxies were classified as disk dominated with an agreement $\geq$80\%.
This suggests that alternative BH fueling methods (i.e., those that do not destroy the disk) are not only more common on the overall AGN population at $z<1$, but also much more efficient than the existing models predict.

\subsubsection{No veiled X-ray activity in merging galaxies}

The aforementioned models leave the possibility that we could be missing an important fraction of AGN due to gas and dust obscuration when a gas-rich major merger is taking place.
Even though obscured AGN can still be detected through their hard X-ray emission \citep{hopkins05_qsoevo}, it is possible that less luminous and highly obscured AGN lie below the detection threshold used to build the X-ray catalogs \citep{treister04}.
The X-ray properties of such obscured objects have been successfully studied in the literature by the means of a stacking analysis of X-ray data \citep[e.g.,][]{daddi07,fiore09}.
If obscured AGN are being missed, they should be preferentially found in merging galaxies.
Therefore, in order to test this scenario and search for this potentially buried X-ray activity, we stack all the inactive galaxies regarded as highly distorted.
Eighty-seven inactive galaxies fulfill our simultaneous criteria of being individually classified as either Dist-1 or Dist-2 with an agreement of $\geq$75\%, and classified as Dist-2 with an agreement $\geq$ 65\%.

For this analysis, we take advantage of the higher sensitivity of the \textit{Chandra} observations of the COSMOS field \citep[C-COSMOS;][]{cosmos_chandra1}, compared with the \textit{XMM-Newton} data.
Even though \textit{Chandra} covered only half of the field ($\sim$0.9 deg$^2$), it has a flux limit three times below the \textit{XMM-Newton} sensitivity, which makes the tradeoff in smaller coverage absolutely justifiable, considering that we want to detect possible X-ray sources below the XMM-COSMOS catalog sensitivity threshold.

\begin{figure}[t]
\centering
\resizebox{\hsize}{!}{\includegraphics{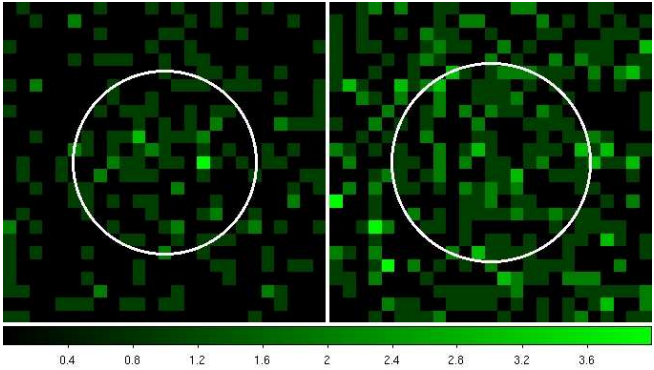}}
\caption{\label{fig_stack}
Stacked \textit{Chandra} images of 45 inactive galaxies likely to be undergoing a major interaction, on the soft 0.5-2 keV (left) and hard 2-8 keV (right) energy bands, showing the average radii of the stacked sources as white circles. The cutouts are $12^{\prime\prime}$ $\times$ $12^{\prime\prime}$.}
\end{figure}

For the stacking of the X-ray data, we used the CSTACK tool\footnote{http://cstack.ucsd.edu} developed by one of the authors (T.M.), which includes a detailed bootstrapping error analysis through 500 realizations.
Because the stacking is made from multiple observations, we consider the counts within a radius varying according to the off-axis angle, corresponding to 90\% of the encircled counts.
We stacked the 45 objects that lie within the C-COSMOS area, after excluding 1 object that was close to an X-ray source.
We found an excess of soft 0.5--2 keV and hard 2--8 keV count rates from the source region at modest levels of $2.2\sigma$ and $2.4\sigma$, respectively.
Figure~\ref{fig_stack} shows the results of the stacking in the two energy bands, with the average radii of $3\farcs4$ and $3\farcs7$ for comparison, within which no source is noticeable above the background level.

The lack of any obvious source after the stacking suggests that this moderate excess could be in part due to the expected emission from star-forming galaxies, and also from extended source emission (e.g., from a galaxy group).
The possibility that a few sources dominate the overall count rate is unlikely since (1) the shape of the count rate distribution is that of a unimodal Gaussian and (2) no outliers are present.
Therefore, it is doubtful that we are missing a significant fraction of accreting BHs hidden within the population of inactive galaxies undergoing interactions.

\subsubsection{No Enhanced Soft X-ray Absorption in Merging AGN Host Galaxies}

As mentioned before, AGN obscuration due to the surrounding gas and dust during a major merger would affect mainly the soft X-ray energy band, while the hard band would remain unobscured.
If we observe an AGN hosted by a merging galaxy, and this interaction was responsible for the BH activity, we would expect to observe a hard X-ray spectrum from this source.
To trace the obscuration level of our interacting AGN host galaxies, we compute their X-ray hardness ratio (HR).
The HR is defined as
\begin{equation}
HR=(H-S)/(H+S),
\end{equation} 
where H and S stand for the hard (2--10 keV) and soft (0.5--2 keV) counts, respectively.
At our redshift range, it is still safe to say that the HR values lower than --0.2 correspond to an unabsorbed, soft spectrum \citep{hasinger08}.

From our visual analysis, we have 13 AGN host galaxies regarded as highly distorted with high agreement according to the criteria used before.
By computing the HR for these objects, we find that, contrary to what models predict, all of these particular sources present soft X-ray spectra.
All of them have HR values $\leq$--0.2, with a mean of --0.53, which shows a low attenuation in the soft band.

It has been argued, however, that the HR diagnostic is rather crude in terms of predicting obscuration, and indeed, bright Compton-thick AGN can feature soft X-ray spectra due to photoionized gas \citep{levenson06}.
Even so, this is only valid when the AGN is not observed directly, and we can easily establish that at least for the type-1 subsample this would not the case, and that we are certainly looking at active, accreting BHs.
Looking only at the seven type-1 objects from these likely merging galaxies, we find that the average HR is --0.56 which indeed suggests a low level of obscuration.

One possible interpretation is that these interactions are not related to the observed AGN episode and that are instead only chance encounters.
Dissipationless or gas-poor mergers could account for the lack of obscuration, but then it is unlikely that any strong merging signatures and substantial accretion onto the central BH would take place directly due to these kind of events.
\citet{pierce07} found the same result for X-ray-selected AGN hosted by interacting galaxies, suggesting that the observed interactions were not responsible for the fueling of those accreting BHs.

From another perspective, however, the models mentioned earlier are limited by the proposed picture of the merger-ULIRG-feedback-quasar timeline \citep[e.g.,][]{sanders88a, sanders88b, hopkins08_framework}, which is already regarded as oversimplified.
The AGN phase is said to happen after coalescence, but observations of large samples of ULIRGs, all of them undergoing interactions, have found a significant scatter in the trends of AGN contribution, accretion rate, and dust obscuration with merging state \citep{veilleux09}.
Some of these have even been found to be dominated by the AGN in pre-merging state.
Chaotic behavior during a merger event can lead to various unpredictable episodes of starburst and nuclear activity.
Such episodic behavior can start much earlier than the final coalescence and can be responsible for different periods of gas inflows, obscuration, and visibility, therefore explaining an already unobscured merger-induced AGN still early during the interaction, as traced by the soft spectra observed in our interacting active galaxies.
This conclusion, at the same time, contradicts the alternative time lag scenario.

\subsection{Major Merging: Not the Most Relevant Mechanism}

Our analysis has demonstrated that the scenario in which mergers are responsible for triggering AGN after a significant time lag is unlikely.
The high fraction of disks, the lack of a hidden significant AGN signal in merging inactive galaxies, and the missing soft X-ray obscuration of interacting AGN hosts all appear to rule out this model as a possible explanation of our results.
The absence of any further evidence in support of this scenario leads us to the only remaining possible interpretation of our results:
active galaxies are involved in major mergers no more frequently than inactive galaxies, and mergers have not played a leading role in AGN triggering for the last 7.5 Gyr.
Our results agree with the few recent studies that have used a control sample \citep[and also with recent results from the E-CDFS by B\"{o}hm et al. 2010, submitted]{dunlop03, grogin03, grogin05, pierce07, gabor09, reichard09, tal09}, in the sense that the morphologies of the AGN host galaxies are not unusual and do not show a preference for merging systems.
Of the studies mentioned earlier which supported a merger-AGN connection, many only provided circumstantial evidence for such a link, without any control sample comparisons.

The lack of enhancement on merging signatures for AGN hosts with respect to the background level indicates that there is no causal connection between merging and AGN triggering up to $z\sim 1$ and $M_\ast\sim10^{11.7}M_{\odot}$, the galaxies dominating BH growth at these redshifts.
It is still a plausible scenario that major mergers could be responsible for some of the brightest quasars; we do not intend to neglect this possibility, but in the context of a clean, large X-ray selected population of AGN, it is certainly not the most relevant mechanism.
The large fraction of AGN hosted by disk-dominated galaxies shows that alternative mechanisms, i.e., stochastic processes and minor mergers dominate, for this sample of objects.

The merger-starburst connection has also been widely studied in the same perspective.
Both mechanisms share the need for enough cold gas to be brought to the central regions of the galaxy, so it is worth mentioning analogous conclusions from the recent literature:
(1) indeed, major mergers \textit{can} trigger strong starbursts \citep[e.g.,][]{mihos&hernquist96, springel00}, but
(2) \textit{not always}, as seen in models \citep{dimatteo07} and observations \citep{bergvall03}, and
(3) its overall contribution is relatively modest \citep{dimatteo08, jogee09}, with no more than 10\% of star formation in massive galaxies being triggered by major mergers at $z\sim 0.6$ \citep{robaina09}.

Different studies \citep[e.g.,][]{ballantyne06, hasinger08, li10} have converged on proposing the following scenario:
the major merger-driven evolution dominates early in the universe, producing the bulk of the brightest quasars at $z=2-3$.
Around $z\sim 1$ however, a different evolutionary mechanism takes over, with secular processes becoming the main triggers for the BH activity and growth.
While our analysis cannot be performed at higher redshifts with the current observational data set, our results appear to fit this picture.
Nevertheless, the overall relevance of major merging, even in the early universe, has yet to be determined.
Other recent studies suggest that secular processes play a much larger role:
observations of massive star-forming galaxies at $z\sim 2$ have shown that their buildup has been dominated by cold rapid accretion and secular processes \citep{genzel08}, without the need of major mergers.
It has been stated on the basis of dark matter simulations that the likely number of major mergers is insufficient to account for the transformation of star-forming turbulent disks at $z=2$ into ellipticals at $z=0$ \citep{genel08}.
A broader view of the accretion history of dark matter halos by \citet{genel10}, quantified that $\sim$60\% of the dark matter in a given halo is contributed by mergers, with only $\sim$20\% being major mergers.
Instead, the rest ($\sim$40\%) of the dark matter would be accreted smoothly.
This also agrees with recent work using smooth particle hydrodynamic simulations, stating that galaxies have acquired most of their baryonic mass through the cold mode of accretion \citep{keres05, keres09}.
Furthermore, merger-free models have shown that isolated galaxies can reproduce the quasar duty cycles between $z=1$ and 3 and feed their BHs with the recycled gas from evolving stars \citep{ciotti07} and even reproduce the observed scaling relations \citep{lusso10}.
Overall, these studies have shown that secular evolution can be highly relevant, also at the redshifts at which the peak of quasar activity occurs.

\section{Conclusions}

In this work, we performed a consistent visual analysis on the \textit{HST}-based morphologies of a sample of 140 X-ray-selected AGN host galaxies over $z\sim 0.3 - 1.0$ and $M_\ast< 10^{11.7}M_\odot$, and compared them with a matched control sample of inactive galaxies under the same conditions.
Our goal was to search for the presence of any significant connection between major merging and BH fueling as suggested by models and observational tests.
In summary:

\begin{itemize}
\item[1.] From our visual analysis, $\sim$85\% of our AGN host galaxies show no strong distortions on their morphologies. Comparison with the control sample shows that the distortion fractions are equal within the 68\% central confidence level. Given our sample size, we can state that at a 95\% confidence level the highly distorted fraction of AGN hosts is less than 1.9 times that of the inactive galaxies. Mergers and interactions involving AGN hosts are not dominant, and occur no more frequently than for inactive galaxies.
\item[2.] Over 55\% of the AGN from our sample are hosted by disk-dominated galaxies, implying a triggering mechanism that would not destroy the disk, i.e., not a major merger. This also indicates that it is unlikely that we could be missing major mergers due to strong distortions having already been washed out over a large time lag prior to the ignition of the AGN. The presence of an important fraction of disk-dominated hosts on the AGN brighter than $L_X>10^{44}$erg s$^{-1}$ suggests that secular fueling mechanisms can be highly efficient as well.
\item[3.] Through a detailed stacking analysis of the X-ray data of our inactive galaxies undergoing mergers, we did not find an underlying X-ray signal indicating the presence of a substantial population of obscured AGN.
\item[4.] Looking at the hardness of the X-ray emission of our AGN hosts that are clearly undergoing an interaction, we found soft X-ray spectra in all of them, contradicting the expected obscuration in this band predicted by models. This can be either because the observed interactions are not responsible for the BH fueling or the unpredictable output of a merger event allows many accretion phases as well as an unobscured AGN, even during such early stages.
\end{itemize}

Our work explicitly suggests that, at least for the last 7.5 Gyr, major merging has not been the most relevant mechanism in the triggering of typical AGN, and that the bulk of the BH accretion occurs through internal secular processes and minor interactions.
The alternative interpretation of a time lag between merger trigger and AGN onset is unlikely due to the zero enhancement of the distortion fraction, the high incidence of disks, and the absence of a significant X-ray signal in merging inactive galaxies as a potential buried AGN population.


\acknowledgments
M.C. thanks G. De Rosa for productive discussions, C.M. Urry for useful comments, and the anonymous referee for practical suggestions.
M.C., K.J., and K.I. are supported through the Emmy Noether Programme of the German Science Foundation (DFG).
TM acknowledges support by CONACyT Apoyo 83564 and UNAM-DGAPA PAPIIT IN110209.

{\it Facilities:} \facility{{\it CXO}}, \facility{VLT:Melipal (VIMOS)}, \facility{{\it HST} (ACS)},
\facility{Magellan:Baade (IMACS)}, \facility{Subaru (SuprimeCam)}, \facility{{\it XMM}}.




\begin{appendix}

\section{AGN-host galaxy decomposition}

The light distribution of the type-1 AGN is clearly dominated by the bright active nucleus, and because we want to analyze the morphologies of their host galaxies, accurate removal of the nuclear source is of vital importance.
This is done through a rigorous two-dimensional parametric fitting with GALFIT, with which we reduce each system down to a two-component model: a PSF to represent the AGN and a S\'{e}rsic light profile accounting for the host galaxy.
After subtraction of the modeled PSF, we are left with the host galaxy emission plus some residuals.
Previous simulations have shown that, at our resolution and S/N, it is sufficient with a single-component model to account for the host galaxy rather than a more complex, multi-component one \citep{sanchez04, simmons08}.
Regarding double-nucleus sources, which could need a dedicated modeling, we only found 1 object in our sample.
This is also consistent with the visual inspection from \citet{civano10} on C-COSMOS sources.
We checked for this particular source (Figure~\ref{fig_dist}, bottom left), for which our method removes the brightest of the two nuclei.
The other point source is significantly fainter and does not dominate the overall galaxy brightness, hence not requiring a more complex decomposition.

An appropriate initial guess of the parameters is recommended to get a faster and converging model with GALFIT. We opt to run Source Extractor \citep{sextractor} on our cutouts to generate, in a fast and automatic way, rough estimates of the free parameters of the S\'{e}rsic profile, such as coordinates within the image, observed magnitude, axis-ratio $b/a$, half-light radius $R_e$, and position angle.

To ensure a reliable decomposition, we take particular care that the host galaxy is modeled with the least possible unnecessary flux transfer between PSF and the S\'{e}rsic profile, which would result in either an over- or undersubtraction.
We perform several GALFIT runs on each object with 3 different choices for the S\'{e}rsic index $n$: we fix it to a $n=1$ exponential profile \citep{freeman70}, to a $n=4$ \citet{devaucouleurs48} profile, and we also leave it a as free parameter for GALFIT to decide. 
To choose the right model, we need our best fit to be reasonable in terms of the resulting parameters. We require our host galaxy model:
(1) not to be too concentrated or too shallow, meaning a half-light radius between 2.5 pixels $< R_e <$ 100 pixels, (2) not to diverge to extreme elongations, therefore to have $b/a >$ 0.5, and (3) to have its S\'{e}rsic index within 0.5 $< n <$ 8, for the free $n$ case. We interpret that if the values run away from these boundaries, GALFIT did not manage to model the underlying galaxy but instead could be accounting for uncertainties in the PSF.

The model with the least $\chi^2$ is chosen between those that comply with the above criteria.
If the model with the free S\'{e}rsic index is chosen among the 3 cases, we reassign the index and model it as follows: (1) if $n<2$ then $n=1$, (2) if $2\leq n \leq3$ then $n=2.5$, and (3) if $n>3$ then $n=4$.

A key aspect of the AGN-host galaxy decomposition is the choice of an accurate PSF, both for modeling the AGN itself and for deconvolving the host galaxy light distribution. Even though the space-based \textit{HST} provides extremely precise PSFs, instrumental effects are still important. The position of the target within the detector and the temporal variability along different orbits can lead to discrepancies between the PSFs from the observations and the ones used for the analysis.
This yields systematic errors in the image decomposition which can be critical for very bright AGN.
The COSMOS survey provides us with the opportunity to minimize these spatial and temporal effects by using stellar PSFs from stars observed under the same conditions than our targets.
For each object, we build specific PSFs by averaging the nearest $\sim$30 stars in the same manner to other similar studies with large \textit{HST} coverage \citep{jahnke04, sanchez04}.
The rms error from the creation of each PSF is propagated to the intrinsic variance of the AGN; the uncertainty of the object being fitted is required by GALFIT in order to converge to a minimum normalized $\chi^2$.

\begin{figure*}[t]
\centering
\resizebox{0.9\textwidth}{!}{\includegraphics{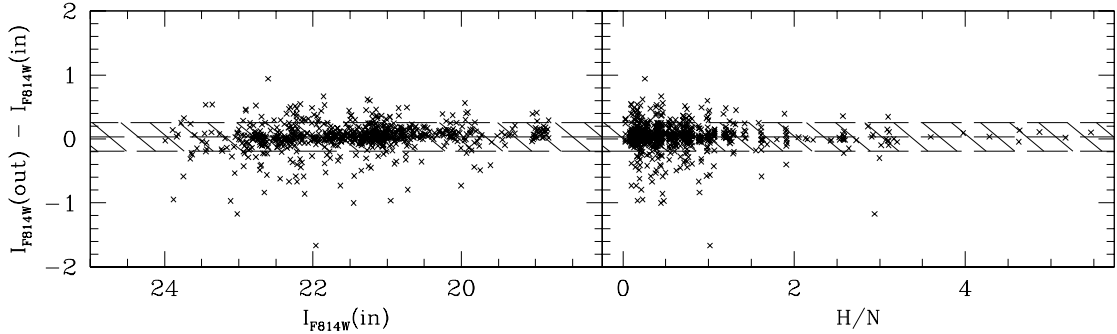}}
\caption{The difference in the observed magnitudes ($I_{F814W}$) of the comparison galaxies before (in) and after (out) the point source addition/subtraction.
The left-hand panel plots this difference against the initial magnitude, and the right-hand panel against the host to nucleus flux ratio, H/N.
The $1\sigma$ deviation away from the mean is 0.23 mag, indicated by the shaded area centered at 0.03 mag.\label{fig_imp}}
\end{figure*}

\section{Creating mock AGN hosts}

For our subsample of type-1 AGN, we build a special comparison sample of simulated AGN hosts by adding stars as fake nuclei to our inactive galaxies.
To remain true to the characteristic blue colors of the AGN, we perform an initial selection of stars from the COSMOS ACS archive by placing color cuts in ($B-V$) $<$ 0.75 and ($V-R$) $<$ 0.95.
For each of the control galaxies, we look for stars that match the contrast level between the fluxes of the host and nucleus (H/N) of the corresponding AGN. With a matching star found, we simply add it over the centroid of the galaxy.

We then apply the same point source removal procedure as for the original type-1 sources.
PSFs are created exactly as before, and the light contributions of the star and the underlying galaxy are separated using GALFIT.
With the exception of 3 unsuccessful fits, we are left with 727 simulated nucleus-subtracted AGN host galaxies that will serve as an appropriate comparison sample for our type-1 AGN hosts.

\section{Testing the reliability of the image decomposition}

The creation of a sample of simulated nucleus-subtracted hosts from a starting point of real galaxies and stars gives us the opportunity to check the impact of our point source removal technique, and to see whether this technique is biased. How significant are the residuals? We have performed photometry on the control galaxies before and after the addition/subtraction of the fake nucleus. If a large magnitude offset were to be found, we would have had to consider reselecting our control sample, because we would inevitably be comparing active and inactive galaxies with different observed magnitudes.
We find that, on average, the galaxies are fainter by 0.03 mag after the subtraction, with a 1$\sigma$ deviation of 0.23 mag.
Figure~\ref{fig_imp} shows the difference between the initial and recovered magnitudes for the hosts as a function of the initial magnitudes and H/N ratio for our control galaxies.
There is no obvious correlation between the offset and the initial magnitudes of the galaxies, but as expected the recovered values tend to be less exact for more compact galaxies and brighter active nuclei.

These results show that this technique is trustworthy, and the offset found can be considered negligible and does not affect our choice of a comparison sample.

\end{appendix}

\twocolumngrid

\begin{thebibliography}{143}
\expandafter\ifx\csname natexlab\endcsname\relax\def\natexlab#1{#1}\fi

\bibitem[{{Bahcall} {et~al.}(1997){Bahcall}, {Kirhakos}, {Saxe}, \&
  {Schneider}}]{bahcall97}
{Bahcall}, J.~N., {Kirhakos}, S., {Saxe}, D.~H., \& {Schneider}, D.~P. 1997,
  \apj, 479, 642

\bibitem[{{Ballantyne} {et~al.}(2006){Ballantyne}, {Everett}, \&
  {Murray}}]{ballantyne06}
{Ballantyne}, D.~R., {Everett}, J.~E., \& {Murray}, N. 2006, \apj, 639, 740

\bibitem[{{Barnes} \& {Hernquist}(1996)}]{barnes&hernquist96}
{Barnes}, J.~E., \& {Hernquist}, L. 1996, \apj, 471, 115

\bibitem[{{Barnes} \& {Hernquist}(1991)}]{barnes&hernquist91}
{Barnes}, J.~E., \& {Hernquist}, L.~E. 1991, \apjl, 370, L65

\bibitem[{{Bell} \& {de Jong}(2001)}]{bdj01}
{Bell}, E.~F., \& {de Jong}, R.~S. 2001, \apj, 550, 212

\bibitem[{{Bell} {et~al.}(2006){Bell}, {Phleps}, {Somerville}, {Wolf}, {Borch},
  \& {Meisenheimer}}]{bell06}
{Bell}, E.~F., {Phleps}, S., {Somerville}, R.~S., {Wolf}, C., {Borch}, A., \&
  {Meisenheimer}, K. 2006, \apj, 652, 270

\bibitem[{{Bennert} {et~al.}(2008){Bennert}, {Canalizo}, {Jungwiert},
  {Stockton}, {Schweizer}, {Peng}, \& {Lacy}}]{bennert08}
{Bennert}, N., {Canalizo}, G., {Jungwiert}, B., {Stockton}, A., {Schweizer},
  F., {Peng}, C.~Y., \& {Lacy}, M. 2008, \apj, 677, 846

\bibitem[{{Bergvall} {et~al.}(2003){Bergvall}, {Laurikainen}, \&
  {Aalto}}]{bergvall03}
{Bergvall}, N., {Laurikainen}, E., \& {Aalto}, S. 2003, \aap, 405, 31

\bibitem[{{Bertin} \& {Arnouts}(1996)}]{sextractor}
{Bertin}, E., \& {Arnouts}, S. 1996, \aaps, 117, 393

\bibitem[{{Brusa} {et~al.}(2007){Brusa}, {Zamorani}, {Comastri}, {Hasinger},
  {Cappelluti}, {Civano}, {Finoguenov}, {Mainieri}, {Salvato}, {Vignali},
  {Elvis}, {Fiore}, {Gilli}, {Impey}, {Lilly}, {Mignoli}, {Silverman}, {Trump},
  {Urry}, {Bender}, {Capak}, {Huchra}, {Kneib}, {Koekemoer}, {Leauthaud},
  {Lehmann}, {Massey}, {Matute}, {McCarthy}, {McCracken}, {Rhodes}, {Scoville},
  {Taniguchi}, \& {Thompson}}]{cosmos_brusa07}
{Brusa}, M., {et~al.} 2007, \apjs, 172, 353

\bibitem[{{Brusa} {et~al.}(2010){Brusa}, {Civano}, {Comastri}, {Miyaji},
  {Salvato}, {Zamorani}, {Cappelluti}, {Fiore}, {Hasinger}, {Mainieri},
  {Merloni}, {Bongiorno}, {Capak}, {Elvis}, {Gilli}, {Hao}, {Jahnke},
  {Koekemoer}, {Ilbert}, {Le Floc'h}, {Lusso}, {Mignoli}, {Schinnerer},
  {Silverman}, {Treister}, {Trump}, {Vignali}, {Zamojski}, {Aldcroft},
  {Aussel}, {Bardelli}, {Bolzonella}, {Cappi}, {Caputi}, {Contini},
  {Finoguenov}, {Fruscione}, {Garilli}, {Impey}, {Iovino}, {Iwasawa},
  {Kampczyk}, {Kartaltepe}, {Kneib}, {Knobel}, {Kovac}, {Lamareille},
  {Leborgne}, {Le Brun}, {Le Fevre}, {Lilly}, {Maier}, {McCracken}, {Pello},
  {Peng}, {Perez-Montero}, {de Ravel}, {Sanders}, {Scodeggio}, {Scoville},
  {Tanaka}, {Taniguchi}, {Tasca}, {de la Torre}, {Tresse}, {Vergani}, \&
  {Zucca}}]{cosmos_brusa10}
---. 2010, \apj, 716, 348

\bibitem[{{Bundy} {et~al.}(2009){Bundy}, {Fukugita}, {Ellis}, {Targett},
  {Belli}, \& {Kodama}}]{bundy09}
{Bundy}, K., {Fukugita}, M., {Ellis}, R.~S., {Targett}, T.~A., {Belli}, S., \&
  {Kodama}, T. 2009, \apj, 697, 1369

\bibitem[{{Bundy} {et~al.}(2010){Bundy}, {Scarlata}, {Carollo}, {Ellis},
  {Drory}, {Hopkins}, {Salvato}, {Leauthaud}, {Koekemoer}, {Murray}, {Ilbert},
  {Oesch}, {Ma}, {Capak}, {Pozzetti}, \& {Scoville}}]{bundy10}
{Bundy}, K., {et~al.} 2010, \apj, 719, 1969

\bibitem[{{Canalizo} {et~al.}(2007){Canalizo}, {Bennert}, {Jungwiert},
  {Stockton}, {Schweizer}, {Lacy}, \& {Peng}}]{canalizo07}
{Canalizo}, G., {Bennert}, N., {Jungwiert}, B., {Stockton}, A., {Schweizer},
  F., {Lacy}, M., \& {Peng}, C. 2007, \apj, 669, 801

\bibitem[{{Canalizo} \& {Stockton}(2000)}]{canalizo&stockton00}
{Canalizo}, G., \& {Stockton}, A. 2000, \aj, 120, 1750

\bibitem[{{Canalizo} \& {Stockton}(2001)}]{canalizo&stockton01}
---. 2001, \apj, 555, 719

\bibitem[{{Capak} {et~al.}(2007){Capak}, {Aussel}, {Ajiki}, {McCracken},
  {Mobasher}, {Scoville}, {Shopbell}, {Taniguchi}, {Thompson}, {Tribiano},
  {Sasaki}, {Blain}, {Brusa}, {Carilli}, {Comastri}, {Carollo}, {Cassata},
  {Colbert}, {Ellis}, {Elvis}, {Giavalisco}, {Green}, {Guzzo}, {Hasinger},
  {Ilbert}, {Impey}, {Jahnke}, {Kartaltepe}, {Kneib}, {Koda}, {Koekemoer},
  {Komiyama}, {Leauthaud}, {Lefevre}, {Lilly}, {Liu}, {Massey}, {Miyazaki},
  {Murayama}, {Nagao}, {Peacock}, {Pickles}, {Porciani}, {Renzini}, {Rhodes},
  {Rich}, {Salvato}, {Sanders}, {Scarlata}, {Schiminovich}, {Schinnerer},
  {Scodeggio}, {Sheth}, {Shioya}, {Tasca}, {Taylor}, {Yan}, \&
  {Zamorani}}]{cosmos_capak07}
{Capak}, P., {et~al.} 2007, \apjs, 172, 99

\bibitem[{{Cappelluti} {et~al.}(2009){Cappelluti}, {Brusa}, {Hasinger},
  {Comastri}, {Zamorani}, {Finoguenov}, {Gilli}, {Puccetti}, {Miyaji},
  {Salvato}, {Vignali}, {Aldcroft}, {B{\"o}hringer}, {Brunner}, {Civano},
  {Elvis}, {Fiore}, {Fruscione}, {Griffiths}, {Guzzo}, {Iovino}, {Koekemoer},
  {Mainieri}, {Scoville}, {Shopbell}, {Silverman}, \& {Urry}}]{cosmos_xmm2}
{Cappelluti}, N., {et~al.} 2009, \aap, 497, 635

\bibitem[{{Ciotti} \& {Ostriker}(2007)}]{ciotti07}
{Ciotti}, L., \& {Ostriker}, J.~P. 2007, \apj, 665, 1038

\bibitem[{{Civano} {et~al.}(2010){Civano}, {Elvis}, {Lanzuisi}, {Jahnke},
  {Zamorani}, {Blecha}, {Bongiorno}, {Brusa}, {Comastri}, {Hao}, {Leauthaud},
  {Loeb}, {Mainieri}, {Piconcelli}, {Salvato}, {Scoville}, {Trump}, {Vignali},
  {Aldcroft}, {Bolzonella}, {Bressert}, {Finoguenov}, {Fruscione}, {Koekemoer},
  {Cappelluti}, {Fiore}, {Giodini}, {Gilli}, {Impey}, {Lilly}, {Lusso},
  {Puccetti}, {Silverman}, {Aussel}, {Capak}, {Frayer}, {Le Floch},
  {McCracken}, {Sanders}, {Schiminovich}, \& {Taniguchi}}]{civano10}
{Civano}, F., {et~al.} 2010, \apj, 717, 209

\bibitem[{{Cole} {et~al.}(2000){Cole}, {Lacey}, {Baugh}, \& {Frenk}}]{cole00}
{Cole}, S., {Lacey}, C.~G., {Baugh}, C.~M., \& {Frenk}, C.~S. 2000, \mnras,
  319, 168

\bibitem[{{Cox} {et~al.}(2006){Cox}, {Jonsson}, {Primack}, \&
  {Somerville}}]{cox06}
{Cox}, T.~J., {Jonsson}, P., {Primack}, J.~R., \& {Somerville}, R.~S. 2006,
  \mnras, 373, 1013

\bibitem[{{Cox} {et~al.}(2008){Cox}, {Jonsson}, {Somerville}, {Primack}, \&
  {Dekel}}]{cox08}
{Cox}, T.~J., {Jonsson}, P., {Somerville}, R.~S., {Primack}, J.~R., \& {Dekel},
  A. 2008, \mnras, 384, 386

\bibitem[{{Daddi} {et~al.}(2007){Daddi}, {Alexander}, {Dickinson}, {Gilli},
  {Renzini}, {Elbaz}, {Cimatti}, {Chary}, {Frayer}, {Bauer}, {Brandt},
  {Giavalisco}, {Grogin}, {Huynh}, {Kurk}, {Mignoli}, {Morrison}, {Pope}, \&
  {Ravindranath}}]{daddi07}
{Daddi}, E., {et~al.} 2007, \apj, 670, 173

\bibitem[{{de Vaucouleurs}(1948)}]{devaucouleurs48}
{de Vaucouleurs}, G. 1948, Annales d'Astrophysique, 11, 247

\bibitem[{{Di Matteo} {et~al.}(2008){Di Matteo}, {Bournaud}, {Martig},
  {Combes}, {Melchior}, \& {Semelin}}]{dimatteo08}
{Di Matteo}, P., {Bournaud}, F., {Martig}, M., {Combes}, F., {Melchior}, A., \&
  {Semelin}, B. 2008, \aap, 492, 31

\bibitem[{{Di Matteo} {et~al.}(2007){Di Matteo}, {Combes}, {Melchior}, \&
  {Semelin}}]{dimatteo07}
{Di Matteo}, P., {Combes}, F., {Melchior}, A., \& {Semelin}, B. 2007, \aap,
  468, 61

\bibitem[{{Di Matteo} {et~al.}(2005){Di Matteo}, {Springel}, \&
  {Hernquist}}]{dimatteo05}
{Di Matteo}, T., {Springel}, V., \& {Hernquist}, L. 2005, \nat, 433, 604

\bibitem[{{Dunlop} {et~al.}(2003){Dunlop}, {McLure}, {Kukula}, {Baum}, {O'Dea},
  \& {Hughes}}]{dunlop03}
{Dunlop}, J.~S., {McLure}, R.~J., {Kukula}, M.~J., {Baum}, S.~A., {O'Dea},
  C.~P., \& {Hughes}, D.~H. 2003, \mnras, 340, 1095

\bibitem[{{Elvis} {et~al.}(2009){Elvis}, {Civano}, {Vignali}, {Puccetti},
  {Fiore}, {Cappelluti}, {Aldcroft}, {Fruscione}, {Zamorani}, {Comastri},
  {Brusa}, {Gilli}, {Miyaji}, {Damiani}, {Koekemoer}, {Finoguenov}, {Brunner},
  {Urry}, {Silverman}, {Mainieri}, {Hasinger}, {Griffiths}, {Carollo}, {Hao},
  {Guzzo}, {Blain}, {Calzetti}, {Carilli}, {Capak}, {Ettori}, {Fabbiano},
  {Impey}, {Lilly}, {Mobasher}, {Rich}, {Salvato}, {Sanders}, {Schinnerer},
  {Scoville}, {Shopbell}, {Taylor}, {Taniguchi}, \&
  {Volonteri}}]{cosmos_chandra1}
{Elvis}, M., {et~al.} 2009, \apjs, 184, 158

\bibitem[{{Ferrarese} \& {Merritt}(2000)}]{mbh_sigma2}
{Ferrarese}, L., \& {Merritt}, D. 2000, \apjl, 539, L9

\bibitem[{{Fiore} {et~al.}(2009){Fiore}, {Puccetti}, {Brusa}, {Salvato},
  {Zamorani}, {Aldcroft}, {Aussel}, {Brunner}, {Capak}, {Cappelluti}, {Civano},
  {Comastri}, {Elvis}, {Feruglio}, {Finoguenov}, {Fruscione}, {Gilli},
  {Hasinger}, {Koekemoer}, {Kartaltepe}, {Ilbert}, {Impey}, {Le Floc'h},
  {Lilly}, {Mainieri}, {Martinez-Sansigre}, {McCracken}, {Menci}, {Merloni},
  {Miyaji}, {Sanders}, {Sargent}, {Schinnerer}, {Scoville}, {Silverman},
  {Smolcic}, {Steffen}, {Santini}, {Taniguchi}, {Thompson}, {Trump}, {Vignali},
  {Urry}, \& {Yan}}]{fiore09}
{Fiore}, F., {et~al.} 2009, \apj, 693, 447

\bibitem[{{Freeman}(1970)}]{freeman70}
{Freeman}, K.~C. 1970, \apj, 160, 811

\bibitem[{{Gabor} {et~al.}(2009){Gabor}, {Impey}, {Jahnke}, {Simmons}, {Trump},
  {Koekemoer}, {Brusa}, {Cappelluti}, {Schinnerer}, {Smol{\v c}i{\'c}},
  {Salvato}, {Rhodes}, {Mobasher}, {Capak}, {Massey}, {Leauthaud}, \&
  {Scoville}}]{gabor09}
{Gabor}, J.~M., {et~al.} 2009, \apj, 691, 705

\bibitem[{{Gebhardt} {et~al.}(2000){Gebhardt}, {Bender}, {Bower}, {Dressler},
  {Faber}, {Filippenko}, {Green}, {Grillmair}, {Ho}, {Kormendy}, {Lauer},
  {Magorrian}, {Pinkney}, {Richstone}, \& {Tremaine}}]{mbh_sigma1}
{Gebhardt}, K., {et~al.} 2000, \apjl, 539, L13

\bibitem[{{Gehren} {et~al.}(1984){Gehren}, {Fried}, {Wehinger}, \&
  {Wyckoff}}]{gehren84}
{Gehren}, T., {Fried}, J., {Wehinger}, P.~A., \& {Wyckoff}, S. 1984, \apj, 278,
  11

\bibitem[{{Genel} {et~al.}(2010){Genel}, {Bouch{\'e}}, {Naab}, {Sternberg}, \&
  {Genzel}}]{genel10}
{Genel}, S., {Bouch{\'e}}, N., {Naab}, T., {Sternberg}, A., \& {Genzel}, R.
  2010, \apj, 719, 229

\bibitem[{{Genel} {et~al.}(2008){Genel}, {Genzel}, {Bouch{\'e}}, {Sternberg},
  {Naab}, {Schreiber}, {Shapiro}, {Tacconi}, {Lutz}, {Cresci}, {Buschkamp},
  {Davies}, \& {Hicks}}]{genel08}
{Genel}, S., {et~al.} 2008, \apj, 688, 789

\bibitem[{{Genzel} {et~al.}(2008){Genzel}, {Burkert}, {Bouch{\'e}}, {Cresci},
  {F{\"o}rster Schreiber}, {Shapley}, {Shapiro}, {Tacconi}, {Buschkamp},
  {Cimatti}, {Daddi}, {Davies}, {Eisenhauer}, {Erb}, {Genel}, {Gerhard},
  {Hicks}, {Lutz}, {Naab}, {Ott}, {Rabien}, {Renzini}, {Steidel}, {Sternberg},
  \& {Lilly}}]{genzel08}
{Genzel}, R., {et~al.} 2008, \apj, 687, 59

\bibitem[{{Georgakakis} {et~al.}(2009){Georgakakis}, {Coil}, {Laird},
  {Griffith}, {Nandra}, {Lotz}, {Pierce}, {Cooper}, {Newman}, \&
  {Koekemoer}}]{georgakakis09}
{Georgakakis}, A., {et~al.} 2009, \mnras, 397, 623

\bibitem[{{Giavalisco} {et~al.}(2004){Giavalisco}, {Ferguson}, {Koekemoer},
  {Dickinson}, {Alexander}, {Bauer}, {Bergeron}, {Biagetti}, {Brandt},
  {Casertano}, {Cesarsky}, {Chatzichristou}, {Conselice}, {Cristiani}, {Da
  Costa}, {Dahlen}, {de Mello}, {Eisenhardt}, {Erben}, {Fall}, {Fassnacht},
  {Fosbury}, {Fruchter}, {Gardner}, {Grogin}, {Hook}, {Hornschemeier}, {Idzi},
  {Jogee}, {Kretchmer}, {Laidler}, {Lee}, {Livio}, {Lucas}, {Madau},
  {Mobasher}, {Moustakas}, {Nonino}, {Padovani}, {Papovich}, {Park},
  {Ravindranath}, {Renzini}, {Richardson}, {Riess}, {Rosati}, {Schirmer},
  {Schreier}, {Somerville}, {Spinrad}, {Stern}, {Stiavelli}, {Strolger},
  {Urry}, {Vandame}, {Williams}, \& {Wolf}}]{goods}
{Giavalisco}, M., {et~al.} 2004, \apjl, 600, L93

\bibitem[{{Granato} {et~al.}(2004){Granato}, {De Zotti}, {Silva}, {Bressan}, \&
  {Danese}}]{granato04}
{Granato}, G.~L., {De Zotti}, G., {Silva}, L., {Bressan}, A., \& {Danese}, L.
  2004, \apj, 600, 580

\bibitem[{{Grogin} {et~al.}(2003){Grogin}, {Koekemoer}, {Schreier}, {Bergeron},
  {Giacconi}, {Hasinger}, {Kewley}, {Norman}, {Rosati}, {Tozzi}, \&
  {Zirm}}]{grogin03}
{Grogin}, N.~A., {et~al.} 2003, \apj, 595, 685

\bibitem[{{Grogin} {et~al.}(2005){Grogin}, {Conselice}, {Chatzichristou},
  {Alexander}, {Bauer}, {Hornschemeier}, {Jogee}, {Koekemoer}, {Laidler},
  {Livio}, {Lucas}, {Paolillo}, {Ravindranath}, {Schreier}, {Simmons}, \&
  {Urry}}]{grogin05}
---. 2005, \apjl, 627, L97

\bibitem[{{Gunn}(1979)}]{gunn79}
{Gunn}, J.~E. 1979, in Active Galactic Nuclei, 213--225

\bibitem[{{H{\"a}ring} \& {Rix}(2004)}]{mbh_m2}
{H{\"a}ring}, N., \& {Rix}, H. 2004, \apjl, 604, L89

\bibitem[{{Hasinger}(2008)}]{hasinger08}
{Hasinger}, G. 2008, \aap, 490, 905

\bibitem[{{Hasinger} {et~al.}(2007){Hasinger}, {Cappelluti}, {Brunner},
  {Brusa}, {Comastri}, {Elvis}, {Finoguenov}, {Fiore}, {Franceschini}, {Gilli},
  {Griffiths}, {Lehmann}, {Mainieri}, {Matt}, {Matute}, {Miyaji}, {Molendi},
  {Paltani}, {Sanders}, {Scoville}, {Tresse}, {Urry}, {Vettolani}, \&
  {Zamorani}}]{cosmos_xmm1}
{Hasinger}, G., {et~al.} 2007, \apjs, 172, 29

\bibitem[{{Heckman} {et~al.}(1984){Heckman}, {Bothun}, {Balick}, \&
  {Smith}}]{heckman84}
{Heckman}, T.~M., {Bothun}, G.~D., {Balick}, B., \& {Smith}, E.~P. 1984, \aj,
  89, 958

\bibitem[{{Hernquist}(1989)}]{hernquist89}
{Hernquist}, L. 1989, \nat, 340, 687

\bibitem[{{Hopkins} {et~al.}(2009){Hopkins}, {Cox}, {Younger}, \&
  {Hernquist}}]{hopkins09_survive}
{Hopkins}, P.~F., {Cox}, T.~J., {Younger}, J.~D., \& {Hernquist}, L. 2009,
  \apj, 691, 1168

\bibitem[{{Hopkins} \& {Hernquist}(2006)}]{hopkins06_stochastic}
{Hopkins}, P.~F., \& {Hernquist}, L. 2006, \apjs, 166, 1

\bibitem[{{Hopkins} \& {Hernquist}(2009)}]{hopkins09_fueling}
---. 2009, \apj, 694, 599

\bibitem[{{Hopkins} {et~al.}(2005{\natexlab{a}}){Hopkins}, {Hernquist}, {Cox},
  {Di Matteo}, {Martini}, {Robertson}, \& {Springel}}]{hopkins05_qsoevo}
{Hopkins}, P.~F., {Hernquist}, L., {Cox}, T.~J., {Di Matteo}, T., {Martini},
  P., {Robertson}, B., \& {Springel}, V. 2005{\natexlab{a}}, \apj, 630, 705

\bibitem[{{Hopkins} {et~al.}(2008){Hopkins}, {Hernquist}, {Cox}, \& {Kere{\v
  s}}}]{hopkins08_framework}
{Hopkins}, P.~F., {Hernquist}, L., {Cox}, T.~J., \& {Kere{\v s}}, D. 2008,
  \apjs, 175, 356

\bibitem[{{Hopkins} {et~al.}(2007){Hopkins}, {Hernquist}, {Cox}, {Robertson},
  \& {Krause}}]{hopkins07_bhplane}
{Hopkins}, P.~F., {Hernquist}, L., {Cox}, T.~J., {Robertson}, B., \& {Krause},
  E. 2007, \apj, 669, 45

\bibitem[{{Hopkins} {et~al.}(2005{\natexlab{b}}){Hopkins}, {Hernquist},
  {Martini}, {Cox}, {Robertson}, {Di Matteo}, \&
  {Springel}}]{hopkins05_lifeqso}
{Hopkins}, P.~F., {Hernquist}, L., {Martini}, P., {Cox}, T.~J., {Robertson},
  B., {Di Matteo}, T., \& {Springel}, V. 2005{\natexlab{b}}, \apjl, 625, L71

\bibitem[{{Hutchings} {et~al.}(1984){Hutchings}, {Crampton}, {Campbell},
  {Duncan}, \& {Glendenning}}]{hutchings84}
{Hutchings}, J.~B., {Crampton}, D., {Campbell}, B., {Duncan}, D., \&
  {Glendenning}, B. 1984, \apjs, 55, 319

\bibitem[{{Hutchings} {et~al.}(1988){Hutchings}, {Johnson}, \&
  {Pyke}}]{hutchings88}
{Hutchings}, J.~B., {Johnson}, I., \& {Pyke}, R. 1988, \apjs, 66, 361

\bibitem[{{Hutchings} \& {Neff}(1992)}]{hutchings&neff92}
{Hutchings}, J.~B., \& {Neff}, S.~G. 1992, \aj, 104, 1

\bibitem[{{Ilbert} {et~al.}(2009){Ilbert}, {Capak}, {Salvato}, {Aussel},
  {McCracken}, {Sanders}, {Scoville}, {Kartaltepe}, {Arnouts}, {Floc'h},
  {Mobasher}, {Taniguchi}, {Lamareille}, {Leauthaud}, {Sasaki}, {Thompson},
  {Zamojski}, {Zamorani}, {Bardelli}, {Bolzonella}, {Bongiorno}, {Brusa},
  {Caputi}, {Carollo}, {Contini}, {Cook}, {Coppa}, {Cucciati}, {de la Torre},
  {de Ravel}, {Franzetti}, {Garilli}, {Hasinger}, {Iovino}, {Kampczyk},
  {Kneib}, {Knobel}, {Kovac}, {LeBorgne}, {LeBrun}, {F{\`e}vre}, {Lilly},
  {Looper}, {Maier}, {Mainieri}, {Mellier}, {Mignoli}, {Murayama}, {Pell{\`o}},
  {Peng}, {P{\'e}rez-Montero}, {Renzini}, {Ricciardelli}, {Schiminovich},
  {Scodeggio}, {Shioya}, {Silverman}, {Surace}, {Tanaka}, {Tasca}, {Tresse},
  {Vergani}, \& {Zucca}}]{cosmos_ilbert09}
{Ilbert}, O., {et~al.} 2009, \apj, 690, 1236

\bibitem[{{Ilbert} {et~al.}(2010){Ilbert}, {Salvato}, {Le Floc'h}, {Aussel},
  {Capak}, {McCracken}, {Mobasher}, {Kartaltepe}, {Scoville}, {Sanders},
  {Arnouts}, {Bundy}, {Cassata}, {Kneib}, {Koekemoer}, {Le F{\`e}vre}, {Lilly},
  {Surace}, {Taniguchi}, {Tasca}, {Thompson}, {Tresse}, {Zamojski}, {Zamorani},
  \& {Zucca}}]{cosmos_ilbert10}
---. 2010, \apj, 709, 644

\bibitem[{{Jahnke} \& {Maccio}(2010)}]{jahnke&maccio10_astroph}
{Jahnke}, K., \& {Maccio}, A. 2010, \apj, submitted (arXiv:1006.0482)

\bibitem[{{Jahnke} {et~al.}(2004){Jahnke}, {S{\'a}nchez}, {Wisotzki}, {Barden},
  {Beckwith}, {Bell}, {Borch}, {Caldwell}, {H{\"a}ussler}, {Heymans}, {Jogee},
  {McIntosh}, {Meisenheimer}, {Peng}, {Rix}, {Somerville}, \&
  {Wolf}}]{jahnke04}
{Jahnke}, K., {et~al.} 2004, \apj, 614, 568

\bibitem[{{Jogee}(2006)}]{jogee06}
{Jogee}, S. 2006, in Lecture Notes in Physics, Berlin Springer Verlag, Vol.
  693, Physics of Active Galactic Nuclei at all Scales, ed. {D.~Alloin}, 143--+

\bibitem[{{Jogee} {et~al.}(2009){Jogee}, {Miller}, {Penner}, {Skelton},
  {Conselice}, {Somerville}, {Bell}, {Zheng}, {Rix}, {Robaina}, {Barazza},
  {Barden}, {Borch}, {Beckwith}, {Caldwell}, {Peng}, {Heymans}, {McIntosh},
  {H{\"a}u{\ss}ler}, {Jahnke}, {Meisenheimer}, {Sanchez}, {Wisotzki}, {Wolf},
  \& {Papovich}}]{jogee09}
{Jogee}, S., {et~al.} 2009, \apj, 697, 1971

\bibitem[{{Kartaltepe} {et~al.}(2010){Kartaltepe}, {Sanders}, {Le Floc'h},
  {Frayer}, {Aussel}, {Arnouts}, {Ilbert}, {Salvato}, {Scoville}, {Surace},
  {Yan}, {Capak}, {Caputi}, {Carollo}, {Cassata}, {Civano}, {Hasinger},
  {Koekemoer}, {Le F{\`e}vre}, {Lilly}, {Liu}, {McCracken}, {Schinnerer},
  {Smol{\v c}i{\'c}}, {Taniguchi}, {Thompson}, {Trump}, {Baldassare}, \&
  {Fiorenza}}]{kartaltepe10b}
{Kartaltepe}, J.~S., {et~al.} 2010, \apj, 721, 98

\bibitem[{{Kauffmann} \& {Haehnelt}(2000)}]{kauffmann00}
{Kauffmann}, G., \& {Haehnelt}, M. 2000, \mnras, 311, 576

\bibitem[{{Kauffmann} {et~al.}(1993){Kauffmann}, {White}, \&
  {Guiderdoni}}]{kauffmann93}
{Kauffmann}, G., {White}, S.~D.~M., \& {Guiderdoni}, B. 1993, \mnras, 264, 201

\bibitem[{{Kere{\v s}} {et~al.}(2009){Kere{\v s}}, {Katz}, {Fardal},
  {Dav{\'e}}, \& {Weinberg}}]{keres09}
{Kere{\v s}}, D., {Katz}, N., {Fardal}, M., {Dav{\'e}}, R., \& {Weinberg},
  D.~H. 2009, \mnras, 395, 160

\bibitem[{{Kere{\v s}} {et~al.}(2005){Kere{\v s}}, {Katz}, {Weinberg}, \&
  {Dav{\'e}}}]{keres05}
{Kere{\v s}}, D., {Katz}, N., {Weinberg}, D.~H., \& {Dav{\'e}}, R. 2005,
  \mnras, 363, 2

\bibitem[{{Koekemoer} {et~al.}(2007){Koekemoer}, {Aussel}, {Calzetti}, {Capak},
  {Giavalisco}, {Kneib}, {Leauthaud}, {Le F{\`e}vre}, {McCracken}, {Massey},
  {Mobasher}, {Rhodes}, {Scoville}, \& {Shopbell}}]{cosmos_acs}
{Koekemoer}, A.~M., {et~al.} 2007, \apjs, 172, 196

\bibitem[{{Kormendy} \& {Kennicutt}(2004)}]{kormendy&kennicutt04}
{Kormendy}, J., \& {Kennicutt}, Jr., R.~C. 2004, \araa, 42, 603

\bibitem[{{Kormendy} \& {Richstone}(1995)}]{mbh_l1}
{Kormendy}, J., \& {Richstone}, D. 1995, \araa, 33, 581

\bibitem[{{Leauthaud} {et~al.}(2007){Leauthaud}, {Massey}, {Kneib}, {Rhodes},
  {Johnston}, {Capak}, {Heymans}, {Ellis}, {Koekemoer}, {Le F{\`e}vre},
  {Mellier}, {R{\'e}fr{\'e}gier}, {Robin}, {Scoville}, {Tasca}, {Taylor}, \&
  {Van Waerbeke}}]{cosmos_leauthaud07}
{Leauthaud}, A., {et~al.} 2007, \apjs, 172, 219

\bibitem[{{Levenson} {et~al.}(2006){Levenson}, {Heckman}, {Krolik}, {Weaver},
  \& {{\.Z}ycki}}]{levenson06}
{Levenson}, N.~A., {Heckman}, T.~M., {Krolik}, J.~H., {Weaver}, K.~A., \&
  {{\.Z}ycki}, P.~T. 2006, \apj, 648, 111

\bibitem[{{Li} {et~al.}(2010){Li}, {Wang}, {Yuan}, {Hu}, \& {Zhang}}]{li10}
{Li}, Y., {Wang}, J., {Yuan}, Y., {Hu}, C., \& {Zhang}, S. 2010, \apj, 710, 878

\bibitem[{{Lilly} {et~al.}(2007){Lilly}, {Le F{\`e}vre}, {Renzini}, {Zamorani},
  {Scodeggio}, {Contini}, {Carollo}, {Hasinger}, {Kneib}, {Iovino}, {Le Brun},
  {Maier}, {Mainieri}, {Mignoli}, {Silverman}, {Tasca}, {Bolzonella},
  {Bongiorno}, {Bottini}, {Capak}, {Caputi}, {Cimatti}, {Cucciati}, {Daddi},
  {Feldmann}, {Franzetti}, {Garilli}, {Guzzo}, {Ilbert}, {Kampczyk}, {Kovac},
  {Lamareille}, {Leauthaud}, {Borgne}, {McCracken}, {Marinoni}, {Pello},
  {Ricciardelli}, {Scarlata}, {Vergani}, {Sanders}, {Schinnerer}, {Scoville},
  {Taniguchi}, {Arnouts}, {Aussel}, {Bardelli}, {Brusa}, {Cappi}, {Ciliegi},
  {Finoguenov}, {Foucaud}, {Franceschini}, {Halliday}, {Impey}, {Knobel},
  {Koekemoer}, {Kurk}, {Maccagni}, {Maddox}, {Marano}, {Marconi}, {Meneux},
  {Mobasher}, {Moreau}, {Peacock}, {Porciani}, {Pozzetti}, {Scaramella},
  {Schiminovich}, {Shopbell}, {Smail}, {Thompson}, {Tresse}, {Vettolani},
  {Zanichelli}, \& {Zucca}}]{cosmos_lilly07}
{Lilly}, S.~J., {et~al.} 2007, \apjs, 172, 70

\bibitem[{{Lin} {et~al.}(2008){Lin}, {Patton}, {Koo}, {Casteels}, {Conselice},
  {Faber}, {Lotz}, {Willmer}, {Hsieh}, {Chiueh}, {Newman}, {Novak}, {Weiner},
  \& {Cooper}}]{lin08}
{Lin}, L., {et~al.} 2008, \apj, 681, 232

\bibitem[{{Lintott} {et~al.}(2008){Lintott}, {Schawinski}, {Slosar}, {Land},
  {Bamford}, {Thomas}, {Raddick}, {Nichol}, {Szalay}, {Andreescu}, {Murray}, \&
  {Vandenberg}}]{galaxyzoo}
{Lintott}, C.~J., {et~al.} 2008, \mnras, 389, 1179

\bibitem[{{Lotz} {et~al.}(2008){Lotz}, {Jonsson}, {Cox}, \& {Primack}}]{lotz08}
{Lotz}, J.~M., {Jonsson}, P., {Cox}, T.~J., \& {Primack}, J.~R. 2008, \mnras,
  391, 1137

\bibitem[{{Lusso} \& {Ciotti}(2010)}]{lusso10}
{Lusso}, E., \& {Ciotti}, L. 2010, \aap, accepted (arXiv:1009.5292)

\bibitem[{{Lusso} {et~al.}(2010){Lusso}, {Comastri}, {Vignali}, {Zamorani},
  {Brusa}, {Gilli}, {Iwasawa}, {Salvato}, {Civano}, {Elvis}, {Merloni},
  {Bongiorno}, {Trump}, {Koekemoer}, {Schinnerer}, {Le Floc'h}, {Cappelluti},
  {Jahnke}, {Sargent}, {Silverman}, {Mainieri}, {Fiore}, {Bolzonella}, {Le
  F{\`e}vre}, {Garilli}, {Iovino}, {Kneib}, {Lamareille}, {Lilly}, {Mignoli},
  {Scodeggio}, \& {Vergani}}]{lusso10_lx}
{Lusso}, E., {et~al.} 2010, \aap, 512, A34+

\bibitem[{{Lynden-Bell}(1967)}]{lynden-bell67}
{Lynden-Bell}, D. 1967, \mnras, 136, 101

\bibitem[{{Magorrian} {et~al.}(1998){Magorrian}, {Tremaine}, {Richstone},
  {Bender}, {Bower}, {Dressler}, {Faber}, {Gebhardt}, {Green}, {Grillmair},
  {Kormendy}, \& {Lauer}}]{mbh_l2}
{Magorrian}, J., {et~al.} 1998, \aj, 115, 2285

\bibitem[{{Mainieri} {et~al.}(2007){Mainieri}, {Hasinger}, {Cappelluti},
  {Brusa}, {Brunner}, {Civano}, {Comastri}, {Elvis}, {Finoguenov}, {Fiore},
  {Gilli}, {Lehmann}, {Silverman}, {Tasca}, {Vignali}, {Zamorani},
  {Schinnerer}, {Impey}, {Trump}, {Lilly}, {Maier}, {Griffiths}, {Miyaji},
  {Capak}, {Koekemoer}, {Scoville}, {Shopbell}, \& {Taniguchi}}]{mainieri07}
{Mainieri}, V., {et~al.} 2007, \apjs, 172, 368

\bibitem[{{Malkan} {et~al.}(1998){Malkan}, {Gorjian}, \& {Tam}}]{malkan98}
{Malkan}, M.~A., {Gorjian}, V., \& {Tam}, R. 1998, \apjs, 117, 25

\bibitem[{{Marconi} \& {Hunt}(2003)}]{mbh_m1}
{Marconi}, A., \& {Hunt}, L.~K. 2003, \apjl, 589, L21

\bibitem[{{Martini}(2004)}]{martini04}
{Martini}, P. 2004, in IAU Symposium, Vol. 222, The Interplay Among Black
  Holes, Stars and ISM in Galactic Nuclei, ed. {T.~Storchi-Bergmann, L.~C.~Ho,
  \& H.~R.~Schmitt}, 235--241

\bibitem[{{Mihos} \& {Hernquist}(1996)}]{mihos&hernquist96}
{Mihos}, J.~C., \& {Hernquist}, L. 1996, \apj, 464, 641

\bibitem[{{Mushotzky}(2004)}]{mushotzky04}
{Mushotzky}, R. 2004, in Astrophysics and Space Science Library, Vol. 308,
  Supermassive Black Holes in the Distant Universe, ed. {A.~J.~Barger}, 53--+

\bibitem[{{Peng}(2007)}]{peng07}
{Peng}, C.~Y. 2007, \apj, 671, 1098

\bibitem[{{Peng} {et~al.}(2010){Peng}, {Ho}, {Impey}, \& {Rix}}]{galfit10}
{Peng}, C.~Y., {Ho}, L.~C., {Impey}, C.~D., \& {Rix}, H. 2010, \aj, 139, 2097

\bibitem[{{Peng} {et~al.}(2002){Peng}, {Ho}, {Impey}, \& {Rix}}]{galfit02}
{Peng}, C.~Y., {Ho}, L.~C., {Impey}, C.~D., \& {Rix}, H.-W. 2002, \aj, 124, 266

\bibitem[{{Pierce} {et~al.}(2007){Pierce}, {Lotz}, {Laird}, {Lin}, {Nandra},
  {Primack}, {Faber}, {Barmby}, {Park}, {Willner}, {Gwyn}, {Koo}, {Coil},
  {Cooper}, {Georgakakis}, {Koekemoer}, {Noeske}, {Weiner}, \&
  {Willmer}}]{pierce07}
{Pierce}, C.~M., {et~al.} 2007, \apjl, 660, L19

\bibitem[{{Porciani} {et~al.}(2004){Porciani}, {Magliocchetti}, \&
  {Norberg}}]{porciani04}
{Porciani}, C., {Magliocchetti}, M., \& {Norberg}, P. 2004, \mnras, 355, 1010

\bibitem[{{Ramos Almeida} {et~al.}(2010){Ramos Almeida}, {Tadhunter}, {Inskip},
  {Morganti}, {Holt}, \& {Dicken}}]{ramosalmeida10}
{Ramos Almeida}, C., {Tadhunter}, C.~N., {Inskip}, K.~J., {Morganti}, R.,
  {Holt}, J., \& {Dicken}, D. 2010, \mnras, 1609

\bibitem[{{Reichard} {et~al.}(2009){Reichard}, {Heckman}, {Rudnick},
  {Brinchmann}, {Kauffmann}, \& {Wild}}]{reichard09}
{Reichard}, T.~A., {Heckman}, T.~M., {Rudnick}, G., {Brinchmann}, J.,
  {Kauffmann}, G., \& {Wild}, V. 2009, \apj, 691, 1005

\bibitem[{{Richstone} {et~al.}(1998){Richstone}, {Ajhar}, {Bender}, {Bower},
  {Dressler}, {Faber}, {Filippenko}, {Gebhardt}, {Green}, {Ho}, {Kormendy},
  {Lauer}, {Magorrian}, \& {Tremaine}}]{richstone98}
{Richstone}, D., {et~al.} 1998, \nat, 395, A14+

\bibitem[{{Rix} {et~al.}(2004){Rix}, {Barden}, {Beckwith}, {Bell}, {Borch},
  {Caldwell}, {H{\"a}ussler}, {Jahnke}, {Jogee}, {McIntosh}, {Meisenheimer},
  {Peng}, {Sanchez}, {Somerville}, {Wisotzki}, \& {Wolf}}]{gems}
{Rix}, H., {et~al.} 2004, \apjs, 152, 163

\bibitem[{{Robaina} {et~al.}(2010){Robaina}, {Bell}, {van der Wel},
  {Somerville}, {Skelton}, {McIntosh}, {Meisenheimer}, \& {Wolf}}]{robaina10}
{Robaina}, A.~R., {Bell}, E.~F., {van der Wel}, A., {Somerville}, R.~S.,
  {Skelton}, R.~E., {McIntosh}, D.~H., {Meisenheimer}, K., \& {Wolf}, C. 2010,
  \apj, 719, 844

\bibitem[{{Robaina} {et~al.}(2009){Robaina}, {Bell}, {Skelton}, {Mc Intosh},
  {Somerville}, {Zheng}, {Rix}, {Bacon}, {Balogh}, {Barazza}, {Barden},
  {B{\"o}hm}, {Caldwell}, {Gallazzi}, {Gray}, {H{\"a}ussler}, {Heymans},
  {Jahnke}, {Jogee}, {van Kampen}, {Lane}, {Meisenheimer}, {Papovich}, {Peng},
  {S{\'a}nchez}, {Skibba}, {Taylor}, {Wisotzki}, \& {Wolf}}]{robaina09}
{Robaina}, A.~R., {et~al.} 2009, \apj, 704, 324

\bibitem[{{Salvato} {et~al.}(2009){Salvato}, {Hasinger}, {Ilbert}, {Zamorani},
  {Brusa}, {Scoville}, {Rau}, {Capak}, {Arnouts}, {Aussel}, {Bolzonella},
  {Buongiorno}, {Cappelluti}, {Caputi}, {Civano}, {Cook}, {Elvis}, {Gilli},
  {Jahnke}, {Kartaltepe}, {Impey}, {Lamareille}, {LeFloch}, {Lilly},
  {Mainieri}, {McCarthy}, {McCracken}, {Mignoli}, {Mobasher}, {Murayama},
  {Sasaki}, {Sanders}, {Schiminovich}, {Shioya}, {Shopbell}, {Silverman},
  {Smol{\v c}i{\'c}}, {Surace}, {Taniguchi}, {Thompson}, {Trump}, {Urry}, \&
  {Zamojski}}]{cosmos_salvato09}
{Salvato}, M., {et~al.} 2009, \apj, 690, 1250

\bibitem[{{S{\'a}nchez} {et~al.}(2004){S{\'a}nchez}, {Jahnke}, {Wisotzki},
  {McIntosh}, {Bell}, {Barden}, {Beckwith}, {Borch}, {Caldwell},
  {H{\"a}ussler}, {Jogee}, {Meisenheimer}, {Peng}, {Rix}, {Somerville}, \&
  {Wolf}}]{sanchez04}
{S{\'a}nchez}, S.~F., {et~al.} 2004, \apj, 614, 586

\bibitem[{{Sanders} \& {Mirabel}(1996)}]{sanders&mirabel96}
{Sanders}, D.~B., \& {Mirabel}, I.~F. 1996, \araa, 34, 749

\bibitem[{{Sanders} {et~al.}(1988{\natexlab{a}}){Sanders}, {Soifer}, {Elias},
  {Madore}, {Matthews}, {Neugebauer}, \& {Scoville}}]{sanders88a}
{Sanders}, D.~B., {Soifer}, B.~T., {Elias}, J.~H., {Madore}, B.~F., {Matthews},
  K., {Neugebauer}, G., \& {Scoville}, N.~Z. 1988{\natexlab{a}}, \apj, 325, 74

\bibitem[{{Sanders} {et~al.}(1988{\natexlab{b}}){Sanders}, {Soifer}, {Elias},
  {Neugebauer}, \& {Matthews}}]{sanders88b}
{Sanders}, D.~B., {Soifer}, B.~T., {Elias}, J.~H., {Neugebauer}, G., \&
  {Matthews}, K. 1988{\natexlab{b}}, \apjl, 328, L35

\bibitem[{{Scarlata} {et~al.}(2007){Scarlata}, {Carollo}, {Lilly}, {Sargent},
  {Feldmann}, {Kampczyk}, {Porciani}, {Koekemoer}, {Scoville}, {Kneib},
  {Leauthaud}, {Massey}, {Rhodes}, {Tasca}, {Capak}, {Maier}, {McCracken},
  {Mobasher}, {Renzini}, {Taniguchi}, {Thompson}, {Sheth}, {Ajiki}, {Aussel},
  {Murayama}, {Sanders}, {Sasaki}, {Shioya}, \& {Takahashi}}]{zest}
{Scarlata}, C., {et~al.} 2007, \apjs, 172, 406

\bibitem[{{Schade} {et~al.}(2000){Schade}, {Boyle}, \& {Letawsky}}]{schade00}
{Schade}, D.~J., {Boyle}, B.~J., \& {Letawsky}, M. 2000, \mnras, 315, 498

\bibitem[{{Schawinski} {et~al.}(2010){Schawinski}, {Dowlin}, {Thomas}, {Urry},
  \& {Edmondson}}]{schawinski10}
{Schawinski}, K., {Dowlin}, N., {Thomas}, D., {Urry}, C.~M., \& {Edmondson}, E.
  2010, \apjl, 714, L108

\bibitem[{{Scoville} {et~al.}(2007{\natexlab{a}}){Scoville}, {Abraham},
  {Aussel}, {Barnes}, {Benson}, {Blain}, {Calzetti}, {Comastri}, {Capak},
  {Carilli}, {Carlstrom}, {Carollo}, {Colbert}, {Daddi}, {Ellis}, {Elvis},
  {Ewald}, {Fall}, {Franceschini}, {Giavalisco}, {Green}, {Griffiths}, {Guzzo},
  {Hasinger}, {Impey}, {Kneib}, {Koda}, {Koekemoer}, {Lefevre}, {Lilly}, {Liu},
  {McCracken}, {Massey}, {Mellier}, {Miyazaki}, {Mobasher}, {Mould}, {Norman},
  {Refregier}, {Renzini}, {Rhodes}, {Rich}, {Sanders}, {Schiminovich},
  {Schinnerer}, {Scodeggio}, {Sheth}, {Shopbell}, {Taniguchi}, {Tyson}, {Urry},
  {Van Waerbeke}, {Vettolani}, {White}, \& {Yan}}]{cosmos_hst}
{Scoville}, N., {et~al.} 2007{\natexlab{a}}, \apjs, 172, 38

\bibitem[{{Scoville} {et~al.}(2007{\natexlab{b}}){Scoville}, {Aussel}, {Brusa},
  {Capak}, {Carollo}, {Elvis}, {Giavalisco}, {Guzzo}, {Hasinger}, {Impey},
  {Kneib}, {LeFevre}, {Lilly}, {Mobasher}, {Renzini}, {Rich}, {Sanders},
  {Schinnerer}, {Schminovich}, {Shopbell}, {Taniguchi}, \& {Tyson}}]{cosmos}
---. 2007{\natexlab{b}}, \apjs, 172, 1

\bibitem[{{Sersic}(1968)}]{sersic}
{Sersic}, J.~L. 1968, {Atlas de galaxias australes}, ed. J.~L. Sersic

\bibitem[{{Shen} {et~al.}(2007){Shen}, {Strauss}, {Oguri}, {Hennawi}, {Fan},
  {Richards}, {Hall}, {Gunn}, {Schneider}, {Szalay}, {Thakar}, {Vanden Berk},
  {Anderson}, {Bahcall}, {Connolly}, \& {Knapp}}]{shen07}
{Shen}, Y., {et~al.} 2007, \aj, 133, 2222

\bibitem[{{Simkin} {et~al.}(1980){Simkin}, {Su}, \& {Schwarz}}]{simkin80}
{Simkin}, S.~M., {Su}, H.~J., \& {Schwarz}, M.~P. 1980, \apj, 237, 404

\bibitem[{{Simmons} \& {Urry}(2008)}]{simmons08}
{Simmons}, B.~D., \& {Urry}, C.~M. 2008, \apj, 683, 644

\bibitem[{{Soltan}(1982)}]{soltan82}
{Soltan}, A. 1982, \mnras, 200, 115

\bibitem[{{Somerville} {et~al.}(2008){Somerville}, {Hopkins}, {Cox},
  {Robertson}, \& {Hernquist}}]{somerville08}
{Somerville}, R.~S., {Hopkins}, P.~F., {Cox}, T.~J., {Robertson}, B.~E., \&
  {Hernquist}, L. 2008, \mnras, 391, 481

\bibitem[{{Somerville} {et~al.}(2001){Somerville}, {Primack}, \&
  {Faber}}]{somerville01}
{Somerville}, R.~S., {Primack}, J.~R., \& {Faber}, S.~M. 2001, \mnras, 320, 504

\bibitem[{{Springel}(2000)}]{springel00}
{Springel}, V. 2000, \mnras, 312, 859

\bibitem[{{Springel} {et~al.}(2005){Springel}, {Di Matteo}, \&
  {Hernquist}}]{springel05}
{Springel}, V., {Di Matteo}, T., \& {Hernquist}, L. 2005, \mnras, 361, 776

\bibitem[{{Springel} \& {Hernquist}(2005)}]{springel&hernquist05}
{Springel}, V., \& {Hernquist}, L. 2005, \apjl, 622, L9

\bibitem[{{Stockton}(1982)}]{stockton82}
{Stockton}, A. 1982, \apj, 257, 33

\bibitem[{{Stockton} \& {Ridgway}(1991)}]{stockton91}
{Stockton}, A., \& {Ridgway}, S.~E. 1991, \aj, 102, 488

\bibitem[{{Surace} \& {Sanders}(1999)}]{surace99}
{Surace}, J.~A., \& {Sanders}, D.~B. 1999, \apj, 512, 162

\bibitem[{{Surace} {et~al.}(2000){Surace}, {Sanders}, \& {Evans}}]{surace00}
{Surace}, J.~A., {Sanders}, D.~B., \& {Evans}, A.~S. 2000, \apj, 529, 170

\bibitem[{{Surace} {et~al.}(1998){Surace}, {Sanders}, {Vacca}, {Veilleux}, \&
  {Mazzarella}}]{surace98}
{Surace}, J.~A., {Sanders}, D.~B., {Vacca}, W.~D., {Veilleux}, S., \&
  {Mazzarella}, J.~M. 1998, \apj, 492, 116

\bibitem[{{Tal} {et~al.}(2009){Tal}, {van Dokkum}, {Nelan}, \&
  {Bezanson}}]{tal09}
{Tal}, T., {van Dokkum}, P.~G., {Nelan}, J., \& {Bezanson}, R. 2009, \aj, 138,
  1417

\bibitem[{{Taniguchi}(1999)}]{taniguchi99}
{Taniguchi}, Y. 1999, \apj, 524, 65

\bibitem[{{Toomre}(1977)}]{toomre77}
{Toomre}, A. 1977, in Evolution of Galaxies and Stellar Populations, ed.
  {B.~M.~Tinsley \& R.~B.~Larson}, 401

\bibitem[{{Toomre} \& {Toomre}(1972)}]{toomre&toomre72}
{Toomre}, A., \& {Toomre}, J. 1972, \apj, 178, 623

\bibitem[{{Treister} {et~al.}(2004){Treister}, {Urry}, {Chatzichristou},
  {Bauer}, {Alexander}, {Koekemoer}, {Van Duyne}, {Brandt}, {Bergeron},
  {Stern}, {Moustakas}, {Chary}, {Conselice}, {Cristiani}, \&
  {Grogin}}]{treister04}
{Treister}, E., {et~al.} 2004, \apj, 616, 123

\bibitem[{{Tremaine} {et~al.}(2002){Tremaine}, {Gebhardt}, {Bender}, {Bower},
  {Dressler}, {Faber}, {Filippenko}, {Green}, {Grillmair}, {Ho}, {Kormendy},
  {Lauer}, {Magorrian}, {Pinkney}, \& {Richstone}}]{mbh_sigma3}
{Tremaine}, S., {et~al.} 2002, \apj, 574, 740

\bibitem[{{Trump} {et~al.}(2007){Trump}, {Impey}, {McCarthy}, {Elvis},
  {Huchra}, {Brusa}, {Hasinger}, {Schinnerer}, {Capak}, {Lilly}, \&
  {Scoville}}]{cosmos_trump07}
{Trump}, J.~R., {et~al.} 2007, \apjs, 172, 383

\bibitem[{{Trump} {et~al.}(2009){Trump}, {Impey}, {Elvis}, {McCarthy},
  {Huchra}, {Brusa}, {Salvato}, {Capak}, {Cappelluti}, {Civano}, {Comastri},
  {Gabor}, {Hao}, {Hasinger}, {Jahnke}, {Kelly}, {Lilly}, {Schinnerer},
  {Scoville}, \& {Smol{\v c}i{\'c}}}]{cosmos_trump09}
---. 2009, \apj, 696, 1195

\bibitem[{{Urrutia} {et~al.}(2008){Urrutia}, {Lacy}, \& {Becker}}]{urrutia08}
{Urrutia}, T., {Lacy}, M., \& {Becker}, R.~H. 2008, \apj, 674, 80

\bibitem[{{van der Wel} {et~al.}(2009){van der Wel}, {Rix}, {Holden}, {Bell},
  \& {Robaina}}]{vanderwel09}
{van der Wel}, A., {Rix}, H., {Holden}, B.~P., {Bell}, E.~F., \& {Robaina},
  A.~R. 2009, \apjl, 706, L120

\bibitem[{{Veilleux} {et~al.}(2009){Veilleux}, {Rupke}, {Kim}, {Genzel},
  {Sturm}, {Lutz}, {Contursi}, {Schweitzer}, {Tacconi}, {Netzer}, {Sternberg},
  {Mihos}, {Baker}, {Mazzarella}, {Lord}, {Sanders}, {Stockton}, {Joseph}, \&
  {Barnes}}]{veilleux09}
{Veilleux}, S., {et~al.} 2009, \apjs, 182, 628

\bibitem[{{Volonteri} {et~al.}(2003){Volonteri}, {Haardt}, \&
  {Madau}}]{volonteri03}
{Volonteri}, M., {Haardt}, F., \& {Madau}, P. 2003, \apj, 582, 559

\bibitem[{{Wada}(2004)}]{wada04}
{Wada}, K. 2004, Coevolution of Black Holes and Galaxies, 186

\bibitem[{{Wyithe} \& {Loeb}(2003)}]{wyithe03}
{Wyithe}, J.~S.~B., \& {Loeb}, A. 2003, \apj, 595, 614

\bibitem[{{Yu} \& {Tremaine}(2002)}]{yu&tremaine02}
{Yu}, Q., \& {Tremaine}, S. 2002, \mnras, 335, 965

\bibitem[{{Zakamska} {et~al.}(2006){Zakamska}, {Strauss}, {Krolik}, {Ridgway},
  {Schmidt}, {Smith}, {Heckman}, {Schneider}, {Hao}, \&
  {Brinkmann}}]{zakamska06}
{Zakamska}, N.~L., {et~al.} 2006, \aj, 132, 1496

\end{thebibliography}

\end{document}